\documentclass[manuscript,screen]{acmart}

\usepackage{graphicx}
\usepackage{caption}
\usepackage{blindtext}
\usepackage{tcolorbox}
\usepackage[final]{pdfpages}
\usepackage{lipsum,multicol}
\usepackage{xcolor}
\usepackage{tikz}
\usepackage{listings}
\usepackage{enumitem}
\usepackage{hyperref}
\usepackage{amsfonts}
\usepackage{wrapfig}
\usepackage{subcaption} 
\usepackage{adjustbox}
\usepackage{colortbl}
\usepackage{fancybox}
\usepackage{multirow}
\usepackage[normalem]{ulem}
\useunder{\uline}{\ul}{}
\usepackage{subfiles}

\lstdefinestyle{javastyle}{
    language=Java,
    basicstyle=\ttfamily\small,
    keywordstyle=\color{blue}\bfseries,
    commentstyle=\color{gray},
    stringstyle=\color{red},
    numbers=left,
    numberstyle=\tiny,
    stepnumber=1,
    breaklines=true,
    frame=single
}

\newcommand{\etal}{et al.\xspace}

\newcommand\revision[1]{\textcolor{black}{#1}}

\newtcolorbox{boxK}{
    fontupper = \small,
    sharpish corners, 
    boxrule = 0pt,
    toprule = 0pt, 
}

\newcommand{\secref}[1]{Section.~\ref{#1}\xspace}
\newcommand{\figref}[1]{Fig.~\ref{#1}\xspace}
\newcommand{\tabref}[1]{Table~\ref{#1}\xspace}

\newcommand*\circled[1]{\tikz[baseline=(char.base)]{
            \node[shape=circle,draw,inner sep=0.5pt] (char) {#1};}}

\newcommand{\llmsforcode}{\textit{LLM4Code}\xspace}
\newcommand{\llmsforcodes}{\textit{LLMs4Code}\xspace}
\newcommand{\pii}{\textit{PII}\xspace}
\newcommand{\piiattack}{\textit{PII Attack}\xspace}
\newcommand{\piiattacks}{\textit{PII Attacks}\xspace}
\newcommand{\key}{\textit{Key}\xspace}
\newcommand{\name}{\textit{Name}\xspace}
\newcommand{\piiemail}{\textit{Email}\xspace}
\newcommand{\password}{\textit{Password}\xspace}
\newcommand{\username}{\textit{Username}\xspace}
\newcommand{\ip}{\textit{IP Address}\xspace}

\newcommand{\StableCodeThreeB}{\textit{stable-code-3b}\xspace}

\newcommand{\codeLlamaSevenB}{\textit{CodeLlama-7b}\xspace}
\newcommand{\codeLlamaThirteenB}{\textit{CodeLlama-13b}\xspace}

\newcommand{\qwenTwoFiveCoderThreeB}{\textit{Qwen2.5-Coder-3B}\xspace}
\newcommand{\qwenTwoFiveCoderSevenB}{\textit{Qwen2.5-Coder-7B}\xspace}
\newcommand{\qwenTwoFiveCoderFourteenB}{\textit{Qwen2.5-Coder-14B}\xspace}

\newcommand{\docode}{\textit{do$_{code}$}\xspace}
\newcommand{\scm}{\textit{SCM}\xspace}

\newcommand{\ate}{\textit{ATE}\xspace}
\newcommand{\ates}{\textit{ATEs}\xspace}
\newcommand{\correlation}{\textit{Pearson}\xspace}

\usepackage{xcolor}
\usepackage{enumitem}

\usepackage{amsmath,amssymb}

\newcommand{\ie}[0]{i.e.,}
\newcommand{\eg}[0]{e.g.,}

\usepackage{changepage}
\usepackage{framed}

\makeatletter
{\par\unskip\endMakeFramed}
\makeatother

\definecolor{darkblue}{rgb}{0.2, 0.2, 0.2}

\definecolor{formalshade}{rgb}{0.95,0.95,1}
\newenvironment{formal}{%
  \def\FrameCommand{%
    \hspace{1pt}%
    {\color{gray}\vrule width 2pt}%
    {\color{formalshade}\vrule width 4pt}%
    \colorbox{formalshade}%
  }%
  \MakeFramed{\advance\hsize-\width\FrameRestore}%
  \noindent\hspace{-4.55pt}%
  \begin{adjustwidth}{}{7pt}%
  \vspace{2pt}\vspace{2pt}%
}
{%
  \vspace{2pt}\end{adjustwidth}\endMakeFramed%
}

\newenvironment{reviewercomment}[1][]
{
    \vspace{2mm}
    \noindent
    \textbf{Referee Comment {\textbf{\emph{#1}}}:}
    \vspace{1mm}
    
    \noindent\itshape 
}
{
    \upshape
    \vspace{1mm}
}

\newenvironment{overallcomment}[1][]
{
    \vspace{2mm}
    \noindent
    \textbf{Overall Comment:}
    {\textbf{\emph{#1}}}
    \vspace{1mm}
    
    \noindent\itshape 
}
{
    \upshape
    \vspace{1mm}
}

\newenvironment{authorresponse}
{
    \vspace{2mm}
    \noindent
    \textbf{Author Response:}
    \vspace{1mm}
    
    \noindent\upshape 
}
{
    \vspace{1mm}
}

\usepackage{tikz}

\AtBeginDocument{%
  }

\setcopyright{acmlicensed}
\copyrightyear{2018}
\acmYear{2018}
\acmDOI{XXXXXXX.XXXXXXX}
\acmConference[Conference acronym 'XX]{Make sure to enter the correct
  conference title from your rights confirmation email}{June 03--05,
  2018}{Woodstock, NY}
\acmISBN{978-1-4503-XXXX-X/2018/06}

\begin{document}

\title{A Causal Perspective on the Role of Training Dynamics for Interpreting Privacy Risks in Code Models}

\author{Hua Yang}
\email{hyang45@ncsu.edu}
\authornote{Both authors contributed equally to this research.}
\orcid{0009-0000-1091-0495}
\affiliation{%
  \institution{North Carolina State University}
  \city{Raleigh}
  \state{North Carolina}
  \country{USA}
}

\author{Alejandro Velasco}
\email{svelascodimate@wm.edu}
\authornotemark[1]
\orcid{0000-0002-4829-1017}
\affiliation{%
  \institution{William \& Mary}
  \city{Williamsburg}
  \state{Virginia}
  \country{USA}
}

\author{Sen Fang}
\email{sfang9@ncsu.edu}
\orcid{0000-0002-9918-7180}
\affiliation{%
  \institution{North Carolina State University}
  \city{Raleigh}
  \state{North Carolina}
  \country{USA}
}

\author{Bowen Xu}
\email{bxu22@ncsu.edu}
\orcid{0000-0002-1006-8493}
\affiliation{%
  \institution{North Carolina State University}
  \city{Raleigh}
  \state{North Carolina}
  \country{USA}
}

\author{Denys Poshyvanyk}
\email{dposhyvanyk@wm.edu}
\orcid{0000-0002-5626-7586}
\affiliation{%
  \institution{William \& Mary}
  \city{Williamsburg}
  \state{Virginia}
  \country{USA}
}

\renewcommand{\shortauthors}{Yang et al.}

\begin{abstract}
\label{sec:abstract}
Large Language Models for Code (\llmsforcode) have significantly enhanced developer productivity in real-world software engineering. However, their reliance on open-source repositories (e.g., GitHub) introduces severe privacy risks, as these repositories contain abundant Personally Identifiable Information (\pii), such as \key, \username, \piiemail, \password, \name, and \ip. Sensitive \pii can be memorized during training and later resurfaced during inference, leading to critical privacy breaches. Prior studies have shown that commercial code completion models are indeed capable of reproducing sensitive \pii, further exacerbating concerns about the trustworthiness and compliance of \llmsforcode in sensitive industries. Nonetheless, existing work has largely treated \pii as a homogeneous category, overlooking the heterogeneous risks posed by different \pii types.
To fill this gap, we investigate whether different types of \pii vary in their likelihood of being learned and leaked by \llmsforcode. More specifically, we examine whether the training dynamics of \llmsforcode on each type influence its leakage risk at inference time, and whether this relationship is causal. Our methodology consists of four stages: (i) constructing a high-quality dataset with a diverse set of \pii types sourced from real-world software repositories;
(ii) fine-tuning representative \llmsforcode of different scales and architectures to capture cross-family patterns; (iii) computing training dynamics on real \pii data and (iv) formulating a structural causal model to estimate the causal effect of learning difficulty on leakage, validated through multiple refutation strategies.
\revision{Our experimental results demonstrate that leakage risks vary significantly across \pii types and that correlation and causation do not always align. In the correlation analysis, easy-to-learn instances such as \ip leak more often, while hard-to-learn instances such as \key and \password leak less frequently. The causal analysis shows that ease of learning has a clear causal effect on leakage for the \username, the strongest and most consistent effect we observe, while the high leakage of \ip is not matched by a comparable causal effect. Ambiguous instances display more complex effects, reducing leakage in some settings while increasing it in others. This study provides the first causal evidence that leakage risks differ by \pii type, offering actionable insights for designing type-aware and learnability-aware defense mechanisms in \llmsforcode.}
\end{abstract}



\received{20 February 2007}
\received[revised]{12 March 2009}
\received[accepted]{5 June 2009}

\maketitle

\section{Introduction}
\label{sec:introduction}

The future of software engineering is evolving towards a paradigm where tasks such as code completion, code summarization, and code generation are heavily reliant on \textit{Large Language Models for Code (\llmsforcode)}. \llmsforcode have demonstrated strong capabilities in real-world development, significantly improving productivity. However, their success is based on massive training corpora drawn from open-source repositories such as GitHub \cite{github, lozhkov2024starcoder, roziere2023code}.
Although these open-source repositories provide abundant training material, they also introduce serious privacy risks due to the strong memorization abilities of \llmsforcode. GitHub contains large amounts of \textit{Personally Identifiable Information} (\pii), including \textit{API keys}, \textit{usernames}, \textit{email addresses}, \textit{passwords}, \textit{personal names}, and \textit{IP addresses} \cite{li2023starcoder, huang2024your, krause2023pushed}. When incorporated into training, this sensitive information can persist in a model’s parameters and later resurface at inference time \cite{carlini2022quantifying, yang2024unveiling}. 

Recently, this problem has also surfaced in industrial settings. For example, Microsoft’s AI research team inadvertently exposed an additional 38~TB of private data while publishing an open-source training dataset on GitHub~\cite{wiz2023microsoft38tb}. 
Importantly, the security risk of such incidents is not determined by the precision of sensitive data detection, but by the ability to obtain valid candidates at scale. In real-world attack scenarios, false positives are often inexpensive to filter, whereas recall is the primary bottleneck. Attackers can readily adopt a generate-and-verify strategy: for instance, verifying a leaked API or cloud credential typically requires only a trivial API call. 
Such incidents highlight the growing tension between the utility of \llmsforcode and the regulatory landscape. Frameworks such as the General Data Protection Regulation (GDPR) and the Health Insurance Portability and Accountability Act (HIPAA) impose strict requirements on handling \pii, yet prior studies \cite{niu2023codexleaks, huang2024your} confirm that commercial code completion models can reproduce sensitive information. 


A natural response to such risks is to block or remove all \pii during training data preprocessing. While this approach is widely regarded as a necessary mitigation step, it is increasingly clear that strict \pii removal is neither fully effective nor practically attainable. First, recent studies consistently demonstrate that \pii leakage persists even in state-of-the-art code models, indicating that existing anonymization pipelines fail to comprehensively detect and remove \pii. For example, Niu et al.~\cite{niu2023codexleaks} showed that Codex memorizes and leaks real names and contact information, while Huang et al.~\cite{huang2024your} extracted 2,702 hard-coded credentials from GitHub Copilot. More recently, Cheng et al.~\cite{cheng2025effective} reported that across 8,000 crafted prompts, four advanced LLMs (GPT-4o, GPT-4, GPT-3.5, and Claude-3.5) collectively generated 3,912 authentic \pii instances. 
Second, strict and type-agnostic \pii removal is known to degrade model utility~\cite{kuo2025proactive, patil2023can, jang2023knowledge}. This issue is particularly acute for code models, where rule-based \pii detectors suffer from high false-positive rates. As shown in~\tabref{tab:llm_ref}, only $9.09\%$ of pre-checked ``password'' candidates pass our strict sensitivity threshold; even measured directly against human labels, password precision is only $0.45$ (\tabref{tab:recall}). Treating all detected candidates as sensitive and removing them indiscriminately risks breaking syntactic and semantic structure, thereby harming downstream code understanding and generation performance. Since commercial model providers must balance privacy protection with utility and competitiveness, applying a uniform maximum-strength defense to all \pii types is often impractical in real-world settings.

Together, these limitations motivate the need to move beyond treating \pii leakage as a monolithic phenomenon. Instead, addressing privacy risks in \llmsforcode requires explaining how different types of \pii are learned, memorized, and propagated during training, and how such heterogeneity translates into distinct leakage risks. Our work takes a step toward this goal by providing causal evidence that learning dynamics vary systematically across \pii types, offering an empirical foundation for prioritizing protection efforts toward the most consequential forms of \pii leakage.

We address \revision{these gaps} by asking whether different types of \pii vary in their probability of being learned and subsequently leaked by \llmsforcode. More specifically, we examine whether the ease with which a model acquires a given \pii type during training influences its leakage risk during inference, and whether this relationship is causal. We pursue these issues in three connected dimensions: \textit{the relative learning difficulty of different \pii types}, \textit{how this difficulty translates into leakage under attack}, and \textit{whether training dynamics directly shape leakage behavior}.

Our methodology proceeds in four stages. First, during dataset construction, we scan and extract code snippets containing different types of \pii from real-world LLM training corpora. We design an automated collection and refinement workflow that integrates regex-based filters, LLM-assisted judgment, and heuristic rules, and we also conduct human verification to ensure accuracy and reliability. Second, in the model setup, we fine-tune representative \llmsforcode of varying scales and architectures to capture patterns that generalize across families of models. Third, we compute the training dynamics \cite{swayamdipta2020dataset} of \llmsforcode on real \pii data to measure the relative learning difficulty of each type. Finally, we formulate a structural causal model (\scm) to estimate the effect of learning difficulty on leakage and validate robustness through multiple refutation strategies.

\revision{The results reveal that the probability of \pii being leaked depends on both its type and its training dynamics, and that correlation and causation do not always agree. In the correlation analysis, easy-to-learn instances such as \ip leak more often, while hard-to-learn instances such as \key and \password leak less. The causal analysis refines this picture. Ease of learning has a clear causal effect on leakage for the \username, which is the strongest and most consistent effect we observe, whereas the high leakage of \ip is not matched by a comparable causal effect. Ambiguity plays a more complex role, reducing leakage in some settings while increasing it for the \name, \password, and \piiemail in larger models. These findings highlight that the privacy risk in \llmsforcode is not uniform but is shaped by type-specific learning behavior. In summary:}

\begin{itemize}
    \item We construct a multi-type \pii dataset from real-world code, combining automated detection, LLM refinement, and human validation to ensure high reliability. 
    \item We analyze training dynamics across \pii types, characterizing their relative difficulty in terms of confidence and variability.  
    \item We perform systematic \pii attack experiments to link learning dynamics with leakage behavior across model families and sizes.  
    \item We formulate a structural causal model to quantify the causal effect of learning difficulty on leakage and validate the robustness of the results.  
    \item We provide the first causal evidence that the types of PII differ in their leakage risks, offering actionable information on defenses that are type- and learnability-aware in \llmsforcode.  
\end{itemize}

\noindent\textbf{Replication package}. To promote transparency and facilitate reproducibility, we make our artifacts available to the community at: \url{https://github.com/PrivacyAI4Code/causal_pii_td.git}. It includes the implementation details of our \pii dataset construction procedure, training dynamic computation, and causal inference framework. 

\section{Background}
\label{sec:background}

\subsection{\pii in Code Repositories}

Open-source code repositories often contain a large amount of \pii, some of which may have severe consequences if leaked. Developers widely rely on version control systems such as GitHub, to manage their source code. Previous studies show that many developers struggle to properly manage \pii and may unintentionally commit it to repositories \cite{krause2023pushed}. For example, in 2019, Meli et al.~identified more than 100,000 instances of \pii in public GitHub repositories \cite{meli2019bad}. In 2022, Rahman et al.~found that developers often committed \pii even when warned by secret detection tools; their case study revealed that developers classified nearly 50\% of warnings as false positives, and therefore bypassed alerts and uploaded secrets regardless~\cite{rahman2022secret}. In 2023, TruffleHog reported the discovery of 721 live API keys and passwords in a subset of GitHub’s public Pull Requests and Issue comments, demonstrating that developers were directly pasting sensitive values such as passwords and keys into public text fields~\cite{trufflehog2023githubcomments}. In 2024, more than 39 million API keys, credentials and other sensitive secrets were exposed on GitHub, raising considerable concern among both the developer community and global enterprises~\cite{gbhackers2024massivegithubleak}. Therefore, the presence of \pii in open source repositories remains an urgent and unsolved problem, highlighting the need for more related research.

\subsection{Training Dynamics}
\label{trainingdynamic}

Different training samples often exhibit varying levels of learning difficulty for neural networks~\cite{shen2019learning, pmlr-v162-ethayarajh22a, swayamdipta2020dataset}. Training dynamics provides an effective tool for characterizing and diagnosing individual samples within training datasets~\cite{swayamdipta2020dataset}. In general, training dynamics refers to the behavior of a model in specific instances as training progresses. For example, Toneva et al. identified training examples that are frequently misclassified in later epochs despite being correctly classified earlier \cite{toneva2018empirical}. Ethayarajh et al. proposed Pointwise V-Information as a metric to quantify the difficulty of individual instances \cite{pmlr-v162-ethayarajh22a}.  

Building on this line of work, Swayamdipta et al. introduced \textit{Data Maps}, which leverage training dynamics to categorize training data into three distinct regions: an \textit{ambiguous region}, where the model is uncertain about the sample; an \textit{easy-to-learn region}, which contains the majority of samples that the model can learn reliably; and a \textit{hard-to-learn region}, consisting of samples that the model struggles to learn \cite{swayamdipta2020dataset}.  
Inspired by this methodology, our study employs training dynamics to analyze different types of \pii data in terms of their \textit{confidence} and \textit{variability}.

The \textbf{confidence} of a target PII token is defined as the mean probability that the model assigns to its ground-truth label throughout training epochs $E$:

\[
\hat{\mu}_i = \frac{1}{E} \sum_{e=1}^{E} p_{\theta^{(e)}}(y_i^* \mid x_i), 
\]

where $p_{\theta^{(e)}}$ denotes the probability predicted by the model with parameters $\theta^{(e)}$ at the end of the $e^{th}$ epoch, $y_i^*$ is the true label of the token $i$, and $x_i$ is its context. Intuitively, a high-confidence instance is considered easier for the learner since the model consistently assigns a high probability to the correct label. 

The \textbf{variability} of a target PII token reflects the stability of its confidence over the course of training. It is measured by the standard deviation of the confidence scores of the model in epochs:  

\[
\hat{\sigma}_i = \sqrt{ \frac{1}{E} \sum_{e=1}^{E} \left( p_{\theta^{(e)}}(y_i^* \mid x_i) - \hat{\mu}_i \right)^2 }.
\]

This enables us to determine which types of \pii are easier or harder for \llmsforcode to learn, providing insights that can inform both the memorization of \pii and the design of future defense mechanisms. Lower variability indicates that the model learns the token in a stable manner across epochs, whereas higher variability suggests fluctuations and potential inconsistency in learning.

\revision{Data Maps~\cite{swayamdipta2020dataset} was originally proposed for classification, and its two axes are defined in terms of the probability assigned to the gold label across training epochs. This definition applies to autoregressive code models: next-token prediction in model training is a per-position classification problem over the vocabulary, where the actual next token serves as the gold label. We therefore compute confidence as the mean probability assigned to the gold \pii token across fine-tuning epochs, and variability as the standard deviation of this probability. Prior work has also extended training-dynamics-based data characterization beyond classification: Hanafi et al.~\cite{hanafi2025identifying} extend the Data Maps framework to autoregressive language models.}

\revision{
For \pii values such as \piiemail, \key, \username, and \ip, the exact sensitive value
is often only weakly determined by the surrounding code context, so
persistently high confidence on such tokens can hardly be explained by
generalization alone and is more consistent with instance-specific
memorization~\cite{feldman2020does,carlini2019secret}. Low variability further indicates that this learning behavior is stable across epochs. Moreover, Sablayrolles et al.~\cite{sablayrolles2019white} prove that the Bayes-optimal membership-inference strategy depends only on the per-example loss; since the per-example loss is monotonically related to
the probability assigned to the gold token, this provides a theoretical
motivation for using confidence to characterize memorization propensity
and potential leakage risk. We do not assume that high confidence or low
variability alone implies realized leakage; instead, we use these
training-dynamics features to characterize how different types of \pii are
learned, and then empirically test their relationship with actual
extraction behavior and causal effects in RQ2 and RQ3.
}

\subsection{Causal Interpretability}

To go beyond correlation and rigorously investigate the relationship between learning dynamics and \pii leakage, we adopt a perspective of structural causal modeling (\scm). Causal interpretability, as introduced in \docode \cite{docode}, provides a post hoc framework grounded in Pearl’s ladder of causation \cite{Pearl2018Causality} for distinguishing genuine causal effects from spurious correlations. In the context of privacy for \llmsforcodes, we adapt this framework to analyze how the dynamics of training (\ie whether a \pii instance is easy, ambiguous, or difficult to learn) causally influence the probability of leakage during inference. 

Formally, we frame the analysis as estimating the \textit{causal effect} of different levels of learning difficulty on the outcome of a \piiattack. This allows us to test whether leakage tendencies reflect intrinsic memorization behaviors or superficial correlations. Following \docode, our procedure consists of four steps: \circled{1} \textbf{model} the problem using causal assumptions encoded in a directed acyclic graph representing the relationship between learning dynamics, confounders, and leakage, \circled{2} \textbf{identify} the causal estimand of interest (the effect of learning difficulty on leakage), \circled{3} \textbf{estimate} causal effects using statistical and machine learning methods, and \circled{4} \textbf{validate} these estimates through sensitivity analyses to ensure robustness. This causal framework enables us to quantify whether the observed leakage risks are actually caused by the way \pii is learned during training, rather than being artifacts of data set bias or model variance.

\section{Related Work}

\subsection{\pii Attacks for \llmsforcode}

GitHub-hosted code is often used to train \llmsforcode, whose strong memorization raises risks of PII leakage. Despite GDPR and HIPAA requiring data protection, commercial LLM4Code rarely reveal how they handle PII. Once memorized, PII can persist in these models even if removed from the original repositories.  
Carlini et al.~demonstrated this risk by extracting hundreds of verbatim training sequences, including \pii in the training set, from GPT-2~\cite{carlini2021extracting}. Niu et al.~developed a semi-automated pipeline to extract sensitive personal information from Codex, the model underlying GitHub Copilot, and showed that 8\% of their prompts successfully elicited sensitive data~\cite{niu2023codexleaks}. Lukas et al.~pointed out that data-scrubbing techniques cannot fully prevent \pii leakage, as they must balance the trade-off between minimizing disclosure and preserving dataset utility, and in practice, scrubbing is often imperfect~\cite{lukas2023analyzing}. More recently, Huang et al.~(2024) constructed prompts from GitHub code files containing credentials and demonstrated that \llmsforcode can reproduce precise training data, including commercial systems such as Gemini and ChatGPT~\cite{huang2024your}.

However, these works have not examined the problem from the training perspective, \ie whether the success of a \pii attack is related to what \pii is being targeted. For example, it remains crucial to study the real-world prevalence of different \pii types in code repositories and how they are memorized during training, in order to determine whether distinct \pii types exhibit significant differences in learnability and leakage risk. Such findings can directly inform the design of more effective defense and mitigation strategies.

\subsection{Safeguarding \pii in \llmsforcode}

Protecting \pii from leakage is a necessary technique for the secure deployment of \llmsforcode, yet current defenses against \pii attacks remain limited. Prior studies have explored several directions: for example, Microsoft Presidio \cite{presidio} has been used to anonymize \pii data before training; Hintersdorf et al.~\cite{hintersdorf2024finding} proposed disabling neurons responsible for memorization in diffusion models to mitigate memorization risks; Geng et al.~\cite{geng2025mitigating} introduced advanced unlearning algorithms that can reduce privacy risks in LLMs4Code while preserving their code generation capabilities; and Guo et al.~\cite{guo2025patch} proposed pairing \pii attack frameworks with \llmsforcode as adversarial dual models, leveraging an improved Group Relative Policy Optimization process to realign the models and enhance their robustness against \pii extraction attacks.

While effective in reducing leakage under specific settings, these methods share a common shortcoming: they mitigate observed leakage without explaining the underlying mechanisms that cause different \pii types to leak in the first place. In particular, existing defenses lack interpretability and predictive power, making it difficult to anticipate their effectiveness across \pii types, model architectures, or training regimes. This limitation motivates the need for a causality-based framework that moves beyond correlation and explicitly disentangles the causal relationships between \pii types and \pii leakage. In this work, we adopt a structural causal modeling approach to identify which training signals of each \pii type causally drive leakage risks, thereby providing mechanistic insights that can inform principled and transferable privacy defenses.

\section{Study Design}

\label{sec:exp_settings}

\begin{figure}[t]
  \centering
  \begin{minipage}[t]{0.49\textwidth}
    \centering
    \includegraphics[width=\linewidth]{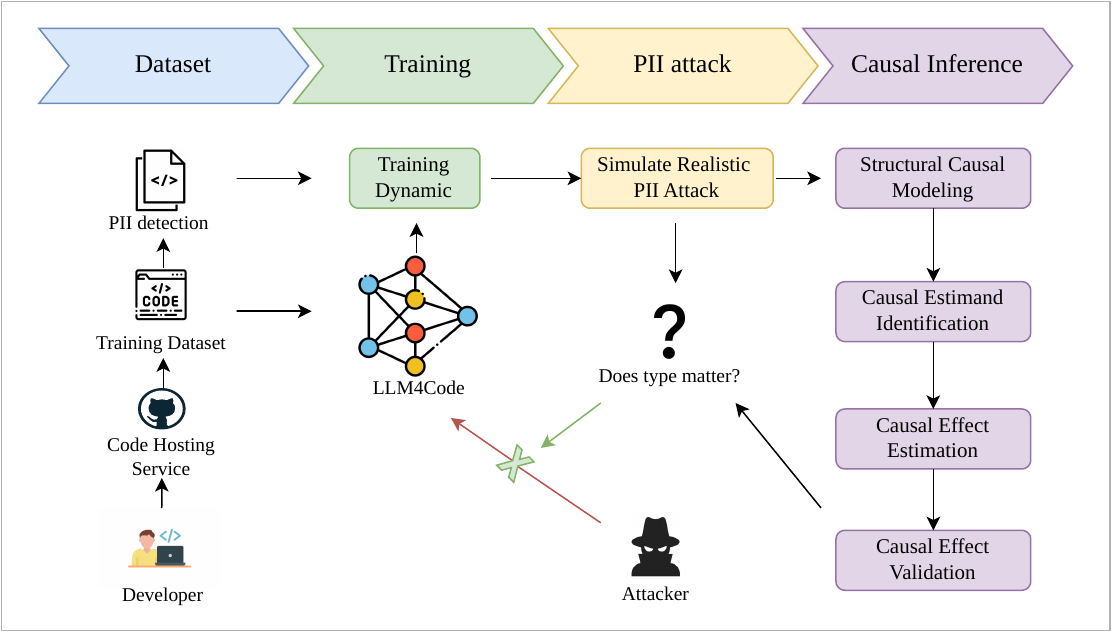}
    \vspace{-2mm}
    \caption{Overview of our study workflow. The process starts with \pii dataset construction, followed by collection of training dynamics during fine-tuning, execution of the \pii attack, and finally causal analysis of the link between training dynamics and attack success rate.}
    \label{fig:pipeline}
  \end{minipage}
  \hfill
  \begin{minipage}[t]{0.49\textwidth}
    \centering
    \includegraphics[width=\linewidth]{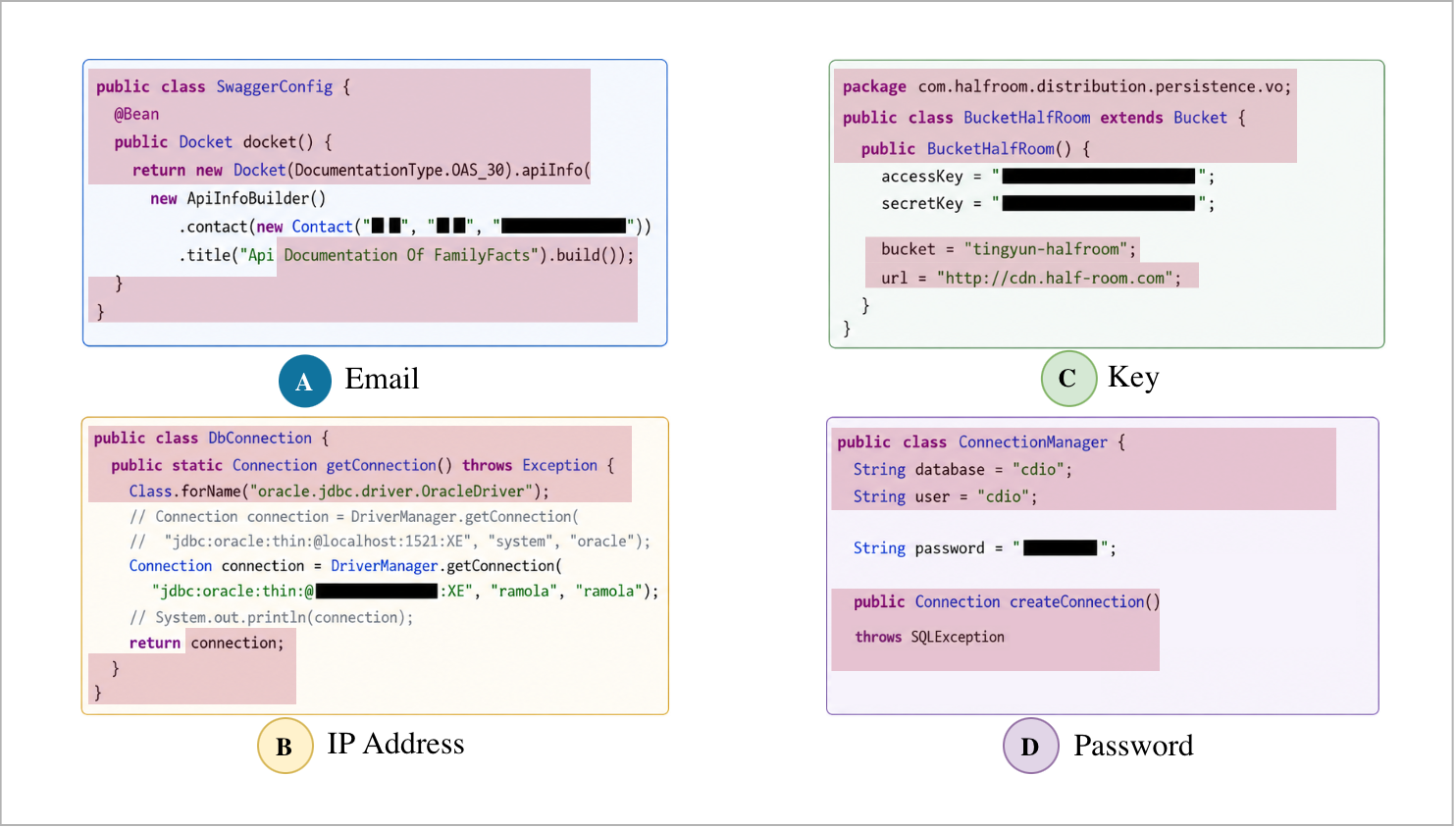}
    \vspace{-2mm}
    \caption{Examples where real \pii values are leaked when the \llmsforcode is queried with surrounding context. Leaked values are redacted with black boxes and query inputs are highlighted in red.}
    \label{fig:example}
  \end{minipage}
  \vspace{-1.5em}
\end{figure}



This section outlines the design of our study. Our experimental workflow is illustrated in \figref{fig:pipeline}. The procedure begins with the construction of the dataset, where we match, refine, and manually verify the code snippets containing different types of \pii. Next, we capture the training dynamics of \pii during the fine-tuning of \llmsforcode using the constructed dataset. We then perform a \pii attack on the fine-tuned \llmsforcode and analyze how the training dynamics of each \pii type relate to the attack success rate. Finally, we adopt causal interpretability by formulating a structural causal model that quantifies the effect of learning difficulty on \pii leakages.

\subsection{Threat Model}
\label{sec:threat_model}
A well-defined threat model that specifies the attacker’s knowledge, assumptions, and capabilities is essential for any rigorous security analysis. To understand the relationship between the type of \pii and \pii attacks, we conduct a realistic \pii attack under the following threat model.

\noindent\textbf{Adversary's objective}. We consider an adversary whose goal is to extract \pii from a code completion model by constructing inputs that partially reproduce the surrounding context of \pii values from open-source repositories, such as GitHub. This assumption is realistic, since training large-scale code generation models without open-source code is virtually impractical. We can always find contextual information accompanying sensitive \pii from open source repositories~\cite{niu2023codexleaks, huang2024your}.


\noindent\textbf{Adversary's capability}. We assume that the attacker has only input–output access to the model. Specifically, the adversary can query the model and observe its generated completions, including the next predicted token. However, the attacker has no access to the internal structure, parameters, or training data of the model. This setting reflects realistic deployment scenarios both commercial and open source \llmsforcodes. For example, commercial models such as OpenAI's GPT-4~\cite{achiam2023gpt} expose only limited information via APIs, while open-source models such as StarCoder~\cite{li2023starcoder} and CodeLlama~\cite{roziere2023code} are often evaluated under similar query constraints to simulate practical audit conditions. We provide actual examples in~\figref{fig:example}. 


\subsection{Research Questions}

Building on the threat model and the study design, we frame our research around three central questions. We progressively move from characterizing the learning behavior of different types of \pii, to examining how such behavior translates into practical leakage risks, and finally to establishing whether a causal relationship exists between the two.

\begin{enumerate}[label=\textbf{RQ$_{\arabic*}$}, ref=\textbf{RQ$_{\arabic*}$}, wide=0pt, labelindent=5pt]\setlength{\itemsep}{0.2em}
    \item \label{rq:learning_difficulty} {\textbf{[Learning Difficulty Across PII Types]} Do different types of \pii exhibit varying degrees of learning difficulty during model training?}\\
    \noindent\textbf{Motivation}: Previous work has shown that \llmsforcode memorize sensitive information from their training corpora \cite{yang2024unveiling, carlini2021extracting, huang2024your, niu2023codexleaks}. However, most studies treat \pii as a single, undifferentiated category without considering the heterogeneity among different types. In reality, different \pii types (\eg \key, \name, and \password) have different formats, distributions, and contextual usage in source code. These differences may influence the ease with which a \llmsforcode learns to memorize them. Understanding whether certain \pii types are inherently easier or harder to learn during training provides critical insights into the underlying risks of memorization.

    \item \label{rq:leakage_relation} {\textbf{[Learning Difficulty–Leakage Risk Relationship]} Does the learning difficulty of different \pii types affect their likelihood of being leaked during attack?}\\
    \noindent\textbf{Motivation}: One might expect that tokens that are easier to learn are more likely to be consistently memorized by \llmsforcode and consequently more vulnerable to attacks, whereas tokens that are harder to learn are less likely to appear in the model's output. However, it remains an open question whether this phenomenon holds across all \pii types. Although different types exhibit varying levels of learning difficulty, it is still unclear whether such differences translate into practical leakage risks. To address this question, we investigate the relationship between the training dynamics of each type and their corresponding attack success rates, thereby linking the memorization process during training to observable leakage behavior inference.

    \item \label{rq:causal_analysis} {\textbf{[Causal Effect of Learning Dynamics on Leakage Risk]} What is the causal relationship between the learning dynamics of different \pii types and their likelihood of being leaked during inference?}\\
    \noindent\textbf{Motivation}: Although \ref{rq:leakage_relation} examines the relationship between learning difficulty and leakage tendency empirically, they do not establish whether there is a causal link between the two. To go beyond the observed correlation, we rigorously investigate whether the way in which a \pii instance is learned directly influences its leakage risk by causal inference. Specifically, we categorize the training dynamics into three levels of learning difficulty: \emph{easy}, \emph{ambiguous}, and \emph{hard}, based on each model's confidence in predicting the true class and the variability of this confidence across epochs. We then compute the causal effect of each difficulty level on the attack success rate for every \pii type, providing a unified view of how the dynamics of memorization shape the privacy risks.
    
\end{enumerate}

\section{Dataset Construction}
\label{sec:dataset}


Although research on software privacy breaches develops datasets containing \pii, such datasets (\eg~\cite{niu2023codexleaks,basak2023secretbench}) are rarely made publicly available due to regulations such as GDPR and HIPAA and to avoid introducing practical security and privacy concerns. Moreover, the contextual distribution of different types of \pii is often difficult to simulate. 
Hence, the synthetic datasets used in the previous work~\cite{he2025privacyxray} are not suitable for our purpose.


\begin{figure}[] 
  \centering
  \includegraphics[width=1\columnwidth]{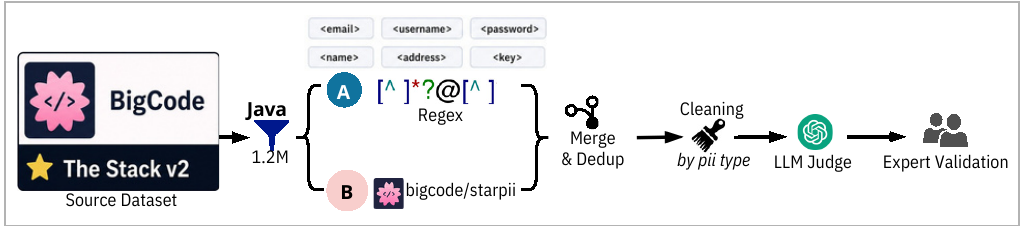}
  \caption{\revision{Dataset Creation Process}}
  \label{fig:dataset_curation}
\end{figure}

To investigate how different types of real-world \pii can be memorized and leaked by \llmsforcode, we constructed a dataset containing six types of \pii from real-world code corpora widely used for training models. The dataset consists of 9,000 code files drawn from the Java subset of The Stack v2, where each file contains at least one sensitive \pii. As a result, each type contains at least 1,500 instances of \pii.

We experimented with multiple detection tools, including TruffleHog~\cite{trufflehog}, Detect-secrets~\cite{detectsecrets}, and Presidio~\cite{presidio}, for \pii matching. However, we found that their performance was unsatisfactory due to various specific limitations (elaborated in \secref{sec:pii_retrieval}). 
Our work thus proposes a methodology for the construction of a type-based \pii dataset, laying the foundation for future research in this direction.


The construction process is illustrated in \figref{fig:dataset_curation}. We designed an automated data collection and refinement pipeline that integrates both rule-based (i.e., regular expressions and heuristics) and LLM-based detection methods. To ensure dataset reliability, two researchers in software engineering conducted a sampling-based annotation test. At a 95\% confidence level with a 5\% margin of error, we estimate that 84.2\% to 94.5\% of the filtered dataset consists of highly plausible sensitive \pii instances.
In the rest of this section, we present the detailed process of how the dataset was constructed, including data collection, \pii retrieval, LLM-based refinement, and expert evaluation.

\subsection{Data Collection}  


We selected a widely used dataset built based on real-world software code, i.e., the \texttt{Stack v2}~\cite{lozhkov2024starcoder}, as our source, ensuring that the identified \pii instances had been used for \llmsforcode training and were highly likely to be memorized. Specifically, we used the \texttt{bigcode/the-stack-v2-dedup} version hosted on Hugging Face, which performs near-deduplication while retaining as many programming languages as possible. This dataset contains 6,457.14 GB of code in 658 different programming languages. To ensure that the fine-tuned models effectively capture the context of \pii, we focus on one of the popular programming languages: Java. We downloaded 1.2M Java files out of the total of 154.28M Java files in \texttt{the-stack-v2-dedup}. From this collection, we constructed a subset of 9,000 highly suspicious \pii instances, demonstrating the widespread presence of \pii in the dataset.  

\subsection{\pii Retrieval} \label{sec:pii_retrieval} 
After experimenting with multiple \pii detection tools, we found that many of them produced excessive false positives, significantly increasing the cost of subsequent processing. Specifically,  we employed three \pii detection tools: TruffleHog~\cite{trufflehog}, Detect-secrets~\cite{detectsecrets}, and Presidio~\cite{presidio}. Detect-secrets only reports the line number where \pii may appear, without providing the actual \pii, making it unsuitable for our task. In our evaluation on 1,000 source files, TruffleHog and Presidio identified 5 and 38,230 \pii matches, respectively. Closer inspection revealed that TruffleHog only flagged five instances of Java Database Connectivity (JDBC) strings, none of which contained actual passwords or production-level public addresses and thus did not constitute real \pii. Presidio, on the other hand, detected 15 different \pii types. However, except for email addresses, the precision of its classifications was below 10\% (based on manual inspection of 50 random samples per type). This limitation comes from the fact that Presidio is designed primarily for processing natural language text, making it not suitable for detecting \pii in the source code.

\begin{table}[t]

\centering
\caption{Statistics of detection results for regex match and StarPII}
\label{tab:search}
\vspace{-1em}

\setlength{\tabcolsep}{6pt} 
\begin{tabular}{ccccccc}
\hline
 &
  \textit{\textbf{Email}} &
  \textit{\textbf{Key}} &
  \textit{\textbf{IP Address}} &
  \textit{\textbf{Name}} &
  \textit{\textbf{Username}} &
  \textit{\textbf{Password}} \\ \hline
\textit{Regex-based PII Detection} &
  53265 &
  601 &
  10554 &
  0 &
  0 &
  0 \\
\textit{StarPII} &
  6137 &
  31855 &
  9608 &
  157751 &
  287461 &
  50564 \\ \hline
\end{tabular}
\end{table}

Through iterative trials and filtering, we employ three complementary methods:

\begin{itemize}
    \item \textbf{Regex-based detection method.} It is worth noting that although we initially used regexes for five \pii types, for accuracy considerations we only retained three types in the final dataset: \piiemail, \key and \ip. The complete set of regexes used for the detection of \pii is provided in our replication package.
    Specifically, the regexes for \piiemail and \ip were adapted from \cite{allal2023santacoder}, while the regex for \key was derived from the list of 18 representative secret types in GitHub Secret Scanning \cite{huang2024your}. The \key type includes widely used credentials issued by major service providers, such as \texttt{aws\_access\_key\_id}. Leakage of such credentials and their corresponding secret key is particularly critical: An attacker could impersonate an AWS identity to perform arbitrary API calls, including downloading S3 data, deleting resources, injecting malicious code, or deploying illicit cryptocurrency mining programs.  
    \item \textbf{LLM-based detection method.} \texttt{StarPII} is a dedicated \pii detection model developed in~\cite{li2023starcoder}. It supports the detection of six types of \pii: \piiemail, \key, \ip, \name, \username, and \password. A detailed breakdown of the retrieval results is presented in \tabref{tab:search}. It should be noted that although regex-based \pii detection identified only 601 instances of \key, the extracted samples exhibit relatively high quality. We merged the results obtained from regex-based detection and \texttt{StarPII}; if two \pii instances from the same code file had identical values, we treated them as the same data point.  
    \item \textbf{Heuristic rules to refine the results}. Following the suggestions of the \texttt{StarPII} developers \cite{huggingface_starpii}, we further applied several heuristic rules to refine the results. For example, we ignored secrets with fewer than 4 characters or more than 300 characters and used the \texttt{ipaddress} Python package to filter out invalid or private IP addresses. For \name, we retained only full names. Despite these refinements, many \pii instances—particularly those in the \password and \key types—remained false positives. Therefore, in the subsequent step, we leveraged LLMs to further refine and validate the extracted \pii data.  
\end{itemize}

\begin{table}[]

\centering
\caption{Statistics of LLM refinement results}
\label{tab:llm_ref}
\vspace{-1em}

\setlength{\tabcolsep}{4pt} 
\begin{tabular}{ccccccc}
\hline
 & \textit{\textbf{Email}} & \textit{\textbf{Key}} & \textit{\textbf{IP Address}} & \textit{\textbf{Name}} & \textit{\textbf{Username}} & \textit{\textbf{Password}} \\ \hline
\textit{Filtered by Pre-check}    & 471     & 4438    & 13388   & 208     & 2104    & 27880  \\
\textit{Evaluated by LLM}         & 2953    & 2356    & 6029    & 2356    & 3425    & 21544  \\
\textit{Exceeding Threshold}      & 2000    & 2000    & 1772    & 2000    & 2000    & 1958   \\
\textit{Exceeding Threshold Rate} & 67.73\% & 84.89\% & 29.39\% & 84.89\% & 58.39\% & 9.09\% \\ \hline
\end{tabular}
\end{table}

\begin{table}[]

\centering
\caption{Detection Performance by Type: Regex vs StarPII vs Combined. R and P denote recall and precision.}
\label{tab:recall}
\begin{tabular}{cccccccc}
\hline
type        & GT   & regex R & regex P & StarPII R & StarPII P & combined R & combined P \\ \hline
email       & 223  & 0.933   & 0.796   & 0.973     & 0.799     & 0.973      & 0.795      \\
password    & 203  & 0.000   & -       & 0.897     & 0.451     & 0.897      & 0.451      \\
name        & 228  & 0.000   & -       & 0.982     & 0.370     & 0.982      & 0.370      \\
username    & 308  & 0.000   & -       & 0.773     & 0.432     & 0.773      & 0.432      \\
ip\_address & 458  & 0.526   & 0.801   & 0.985     & 0.746     & 0.985      & 0.746      \\
key         & 269  & 0.227   & 0.938   & 0.829     & 0.209     & 0.833      & 0.215      \\
overall     & 1689 & 0.302   & 0.813   & 0.909     & 0.438     & 0.909      & 0.446      \\ \hline
\end{tabular}

\end{table}

\revision{Recall and precision of the retrieval stage. A natural concern is whether the retrieval pipeline, while tuned to suppress false positives, misses a substantial fraction of genuine \pii, and whether using regex for three structured types (Email, IP Address, Key) but StarPII for all six introduces type-dependent coverage bias. To assess this directly, we constructed a ground-truth set by having two software-engineering experts (the annotators described in \secref{expert} ) exhaustively label every PII instance in a random sample of files from the Java subset of The Stack v2, rather than stopping at one instance per file. This yields 808 unique files and 1,689 labeled PII spans across the six types (Email 223, Password 203, Name 228, Username 308, IP Address 458, Key 269). We then ran the exact detection stage of our pipeline over this set, removing the one-instance-per-file cap so that the measurement reflects true detector recall rather than the sampling cap; a span is counted as recovered if a detection of the same type overlaps the labeled span. The resulting per-type recall and precision are reported in~\tabref{tab:recall}, and the ground-truth set and evaluation script are included in our replication package.}

\revision{The combined detector recovers 1,536 of 1,689 spans, an overall recall of 0.909, with per-type recall of 0.973 (Email), 0.982 (Name), 0.985 (IP Address), 0.897 (Password), 0.833 (Key), and 0.773 (Username). Two observations address the asymmetry concern. First, regex and StarPII are not parallel retrieval channels: StarPII is applied uniformly to all six types and accounts for nearly all recovered spans, while regex contributes only a single additional span (one Key) beyond what StarPII already finds, so the combined recall is effectively that of StarPII alone and the dominant error profile is shared across types rather than fragmented by detection method. Second, the role of regex is precision rather than recall: its own recall on the structured types is low (0.23 for Key, 0.53 for IP Address), but its precision matches or exceeds StarPII's on all three (Email 0.80 vs.\ 0.80, IP Address 0.80 vs.\ 0.75, Key 0.94 vs.\ 0.21). Because the dataset targets a fixed quota of validated instances per type and refinement is run in precision order, the higher-precision regex candidates are refined first, which reaches the quota with fewer LLM-refinement calls; this is most pronounced for Key, where StarPII precision is only 0.21 and far more candidates would otherwise need refining to collect the same number of genuine instances (\secref{refine_subs}). The two weakest types, Username (0.773) and Key (0.833), together account for 115 of the 153 missed spans; we advise that type-level comparisons involving these two types be read with this residual recall gap in mind.}

\subsection{LLM Refinement} 
\label{refine_subs}

Because manually annotating all detected \pii values would be prohibitively costly in terms of both time and resources, we leveraged LLMs' capability to assess further whether the detected data corresponded to genuine sensitive information. To reduce the cost of LLM usage and accelerate the refinement process, we first designed a set of pre-check rules for each \pii type. These rules filtered out low-quality noisy candidates, which are often caused by formatting errors or placeholder values. The complete list of pre-check rules is presented in \tabref{tab:check} and is included in the replication package. Candidates who did not pass these quick checks were discarded, and only the remaining instances were refined using LLM. For example, for \key, we constructed a list of common test keys and discarded any instances that matched this list. Additionally, we filtered out keys shorter than nine characters or those with Shannon entropy below three.

Through iterative trials and analysis of LLM annotation capabilities, we instructed LLM to evaluate each candidate \pii instance in three dimensions:

\begin{itemize}
\item \textbf{Format}: \textit{What is the valid format for this type of \pii, and what are the common invalid cases?  }

\item \textbf{Context}: \textit{In what contexts is this type of \pii typically sensitive, and does the candidate occur within a test file?}

\item \textbf{Realness}: \textit{Is the candidate a placeholder, a dummy value, or only a partial \pii value?}

\end{itemize}
We incorporated type-specific heuristics to mitigate false positives. For example, for \username, if the string was a nickname or contained only a first name or last name, it was treated as non-sensitive. The complete prompts used in this step are provided in our replication package. We employed \texttt{gpt-5-mini} to refine the collected \pii data using one-shot prompting guided by these criteria. Moreover, after experimenting with context window sizes of 100, 200, and 500 characters, we settled on providing the LLM with 500 preceding and 500 succeeding characters surrounding each target \pii value. The LLM then assigned a sensitivity score between 0 and 100, with higher values indicating a greater likelihood of sensitivity. To ensure balance between \pii types, we enforced a minimum of 1,500 instances per type with sensitivity scores that exceed the threshold. Given that \password and \ip instances exhibited higher false positive rates during subsequent manual validation, we set their thresholds at 95, while using 90 for the other types. We report more details about this process in \tabref{tab:llm_ref}. In particular, only 9.09\% of the \password instances exceeded the threshold, indicating that they are highly prone to false positives.

\subsection{Expert Evaluation}  
\label{expert}

\begin{table}[]

\centering
\caption{Complete list of pre-check rules for each \pii type}
\label{tab:check}
\vspace{-1em}

\begin{tabular}{cl}
\hline
\multicolumn{1}{l}{}     & \multicolumn{1}{c}{\textit{\textbf{Pre-Check Rules}}}                             \\ \hline
                         & Exactly one "@" symbol must exist; both local and domain parts must be non-empty. \\
                         & No leading, trailing, or consecutive (“..”) dots.                                 \\
                         & No whitespace, slashes, or parentheses.                                           \\
\multirow{-4}{*}{\piiemail} &
  \begin{tabular}[c]{@{}l@{}}Domain Validation: The domain must not start or end with a dot. The top-level domain \\ is verified against the official IANA list.\end{tabular} \\
\rowcolor[HTML]{EFEFEF} 
\cellcolor[HTML]{EFEFEF} &
  IP Address must be syntactically valid and informative, i.e., publicly routable. \\
\rowcolor[HTML]{EFEFEF} 
\multirow{-2}{*}{\cellcolor[HTML]{EFEFEF}\ip} &
  \begin{tabular}[c]{@{}l@{}}Private, loopback, link-local, unspecified, multicast, reserved, and broadcast addresses \\ are rejected.\end{tabular} \\
                         & No suspicious characters (slashes, whitespace, braces).                           \\
                         & Minimum length of 16.                                                             \\
                         & At least 8 unique characters.                                                     \\
\multirow{-4}{*}{\key}    & Shannon entropy more than 3.0, indicating sufficient randomness.                  \\
\rowcolor[HTML]{EFEFEF} 
\cellcolor[HTML]{EFEFEF} & Must conform to patterns of full English names (more than 2 parts).               \\
\rowcolor[HTML]{EFEFEF} 
\cellcolor[HTML]{EFEFEF} & Permitted separators: space, period, hyphen, or comma.                            \\
\rowcolor[HTML]{EFEFEF} 
\rowcolor[HTML]{EFEFEF} 
\multirow{-4}{*}{\cellcolor[HTML]{EFEFEF}\name} &
  Strings that fail to match multi-word English name formats are discarded. \\
                         & Minimum length of 8.                                                              \\
                         & All characters must belong to the allowed secure charset.                         \\
\multirow{-3}{*}{\password} &
  \begin{tabular}[c]{@{}l@{}}Exclusion of known weak or common passwords \\ (e.g., “123456”, “admin”, “password1”).\end{tabular} \\
\rowcolor[HTML]{EFEFEF} 
\cellcolor[HTML]{EFEFEF} & Only letters, digits, underscores, and hyphens permitted.                         \\
\rowcolor[HTML]{EFEFEF} 
\cellcolor[HTML]{EFEFEF} & Minimum of 6 characters.                                                          \\
\rowcolor[HTML]{EFEFEF} 
\multirow{-3}{*}{\cellcolor[HTML]{EFEFEF}\username} &
  Disallows known false usernames (\eg “admin”, “root”, “guest”, “testuser”). \\ \hline
\end{tabular}
\setlength{\tabcolsep}{4pt} 

\vspace{-1em}
\end{table}

To validate the reliability of the LLM-refined dataset, we recruited two researchers in software engineering, each with at least four years of Java experience, to manually annotate a subset of the data. \revision{Both annotators have research experience in software security and privacy, including familiarity with all the \pii types studied in this paper. To ensure annotators had sufficient information for their assessments, we first provided them with a brief guideline that explains our research objectives, the definition of each \pii type, and concrete examples in our dataset.}

We first sampled 100 instances of each type to estimate the expected proportion of valid \pii, which was approximately 0.9. To determine the required sample size for manual validation, we applied the standard sample size estimation formula with finite-population correction. For each \pii type, with a population size of 1,500, a 95\% confidence level, and a 5\% margin of error, the minimum required sample size was 127. Consequently, we randomly sampled 150 instances per type, which were independently reviewed by the two experts. Specifically, we asked participants the following question: \textit{``Is the highlighted \pii value in the code file considered sensitive information?''} Participants responded using a three-point Likert scale~\cite{joshi2015likert}, where 1, 2, and 3 corresponded to \textit{Disagree}, \textit{Uncertain}, and \textit{Agree}, respectively. The results showed that for each \pii type, at least 89.33\% of the instances were labeled as 3, indicating that the highlighted values represented sensitive information.

Based on this expert evaluation, we obtained the final dataset consisting of 9,000 code files containing sensitive \pii, evenly distributed over six types with 1,500 instances per type.
We then partitioned the dataset by random sampling into training, validation, and test sets that contain 7,200, 900, and 900 samples, respectively. 

\revision{The validation set is used for model selection and 
hyperparameter tuning, and the test set is held out for final evaluation. Following the 80/10/10 partitioning scheme 
commonly adopted in code analysis research~\cite{chen2023diversevul}, 
the validation and test sets are of equal size (900 samples each), 
ensuring symmetric evaluation conditions.}

\section{Experimental Settings}

To systematically examine how different types of \pii influenced leakage rates, we conducted a comprehensive empirical study. Specifically, we analyzed training dynamics during fine-tuning on our constructed \pii dataset and simulated realistic attack scenarios to evaluate leakage risks. Finally, we employed Structural Causal Modeling (\scm) to control for potential confounders and revealed the causal relationship between training dynamics and attack success rates.

\subsection{Models}

We focus on fine-tuning \llmsforcode, as it has been extensively applied in both research and industry~\cite{he2025privacyxray}. For the downstream task, we select \textit{code completion}, a core feature of nearly all modern integrated development environments and used by prior work in privacy leak research~\cite{huang2024your}. Code completion suggests relevant code snippets based on the current programming context. Analyzing token-level completions allows us to examine how effectively \pii-related tokens are learned and memorized by \llmsforcode. We also select models that support the \textit{Fill-in-the-Middle (FIM)} objective, as the suffix provides additional context that may increase the risk of leakage~\cite{lukas2023analyzing}.

To ensure the generalizability of our findings, we fine-tune models with different architectures and sizes on our constructed dataset and analyze their training dynamics. Specifically, \revision{we include three widely used families of open-source models capable of completing code snippets:}  

\begin{itemize}
    \item \textbf{Stable Code.} Stable Code~\cite{pinnaparaju2024stable} is a decoder-only \llmsforcode trained for code completion that supports FIM. \revision{In our study, we use the \texttt{stabilityai/stable-code-3b} version from Hugging Face.}
    \item \textbf{CodeLlama.} CodeLlama~\cite{roziere2023code} is derived from Llama 2 and designed for code generation and FIM tasks. It supports multiple languages including Python, C++, and Java. We use the models \texttt{codellama/CodeLlama-7b-hf} and \texttt{codellama/CodeLlama-13b-hf}.  
    \item \textbf{Qwen2.5-Coder.} Qwen2.5-Coder~\cite{hui2024qwen2} is based on the Qwen2.5 Transformer architecture and supports long context lengths as well as FIM objectives. It is trained on more than 5.5 trillion tokens from various domains, including code, natural language, mathematics, and synthetic corpora. We use \texttt{Qwen/Qwen2.5-Coder-3B}, \texttt{Qwen/Qwen2.5-Coder-7B}, and \texttt{Qwen/Qwen2.5-Coder-14B}.  
\end{itemize}

In general, our selected models cover three model families and three parameter scales, with at least two families represented at each scale.

\subsection{Evaluation Metrics}  
\label{sec:metrics}


\noindent\textbf{Metrics related to training dynamics}. \textbf{Confidence} measures how strongly a model predicts the correct token on average across training epochs, while \textbf{variability} captures the stability of these predictions (elaborated in \secref{trainingdynamic}). High confidence and low variability indicate that the model consistently learns a token, whereas low confidence or unstable predictions suggest difficulty. To translate these raw signals into interpretable categories, we classify each \pii instance into one of three levels of learning difficulty: \emph{easy-to-learn}, \emph{hard-to-learn}, or \emph{ambiguous}.


\revision{Building on the dataset-cartography methodology of Swayamdipta \etal~\cite{swayamdipta2020dataset}, we categorize each \pii instance from its confidence and variability, both measured across training epochs over the distribution of all \pii instances. Each axis is split at percentiles specified below: confidence at a high and a low percentile, and variability at a single high percentile. Variability is considered first: an instance with high variability is labeled \textbf{ambiguous}, regardless of its confidence, because its predictions keep changing across epochs without settling. Among the remaining low-variability instances, one is labeled \textbf{easy-to-learn} if its confidence is high and \textbf{hard-to-learn} if its confidence is low; in both cases the training behavior is stable and the prediction is, respectively, consistently correct or consistently incorrect. Low-variability instances whose confidence falls in the middle are not clearly separated from either pole, so we exclude them from the easy-versus-hard contrast.}

\revision{The exact cutoff values affect which category each instance falls into. To ensure that our results do not depend on one specific choice, we evaluate two sets of cutoffs with different levels of selectivity. The baseline setting places the easy-to-learn and hard-to-learn categories in the top and bottom quarters of the confidence distribution, using the 75th and 25th percentiles, and identifies ambiguous-to-learn instances using the 75th percentile of variability. These cutoffs are inspired by the original dataset cartography work \cite{swayamdipta2020dataset}. We also evaluate a stricter setting based on the top and bottom fifths, using the 80th and 20th percentiles for confidence and the 80th percentile for variability. This setting retains only the most clearly separated instances. We compute all causal estimates and correlations reported in \secref{sec:rq3_results} under both settings and report the estimate obtained under each setting. This sensitivity analysis shows whether the conclusions remain stable under a more selective definition of learning difficulty.}

\noindent\textbf{Metrics related to vulnerability to \piiattacks}.
Following prior work \cite{huang2024your, niu2023codexleaks}, we evaluate the leakage of fine-tuned models using a black-box query strategy. Specifically, we construct completion inputs consisting of partial prefixes and suffixes of code snippets surrounding a target \pii instance. If the completion of the model contains the target \pii, we count it as a successful attack. Formally, the \textbf{\pii attack success} for a target \pii $i$ is defined as follows. Let the ground-truth \pii token (or token sequence) be $\mathbf{s}_i = (s_{i,1}, \dots, s_{i,m_i})$, and let the query input be constructed from its prefix and suffix context. When the model generates a completion $\hat{\mathbf y}_i = (\hat y_{i,1}, \dots, \hat y_{i,L_t}),$ we define the single-attempt attack success indicator as

\[
A_i^{(t)} = \mathbb{1}\!\left[\, \mathbf{s}_i \sqsubseteq \hat{\mathbf y}_i^{(t)} \,\right],
\]

where $\sqsubseteq$ denotes that $\mathbf{s}_i$ appears as a contiguous subsequence of the completion. A higher attack success rate indicates that the fine-tuned model is more likely to reproduce memorized sensitive \pii tokens when prompted with surrounding context, thereby reflecting greater leakage risk.

\subsection{Structural Causal Modeling}

\begin{figure}[t]
  \centering
  \begin{minipage}{0.49\textwidth}
    \centering
    \includegraphics[width=\linewidth]{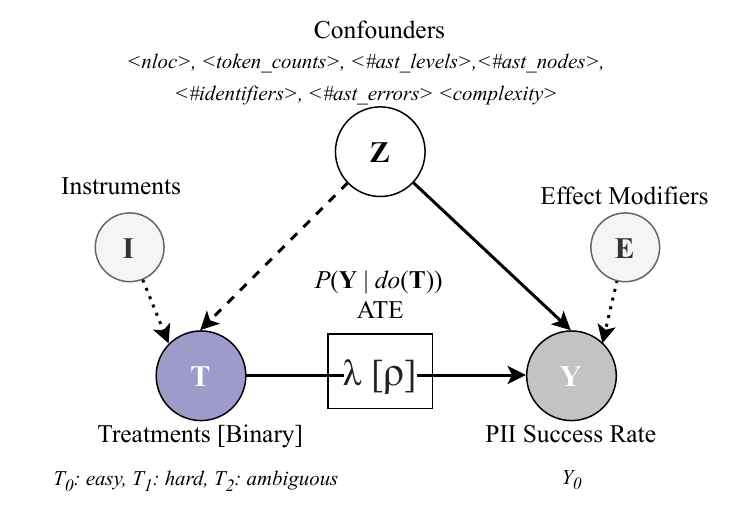}
    \caption{Structural causal model of the effect of learning difficulty on
    \pii leakage.}
    \label{fig:scm}
  \end{minipage}
  \hfill
  \begin{minipage}{0.49\textwidth}
    \centering
    \includegraphics[width=\linewidth]{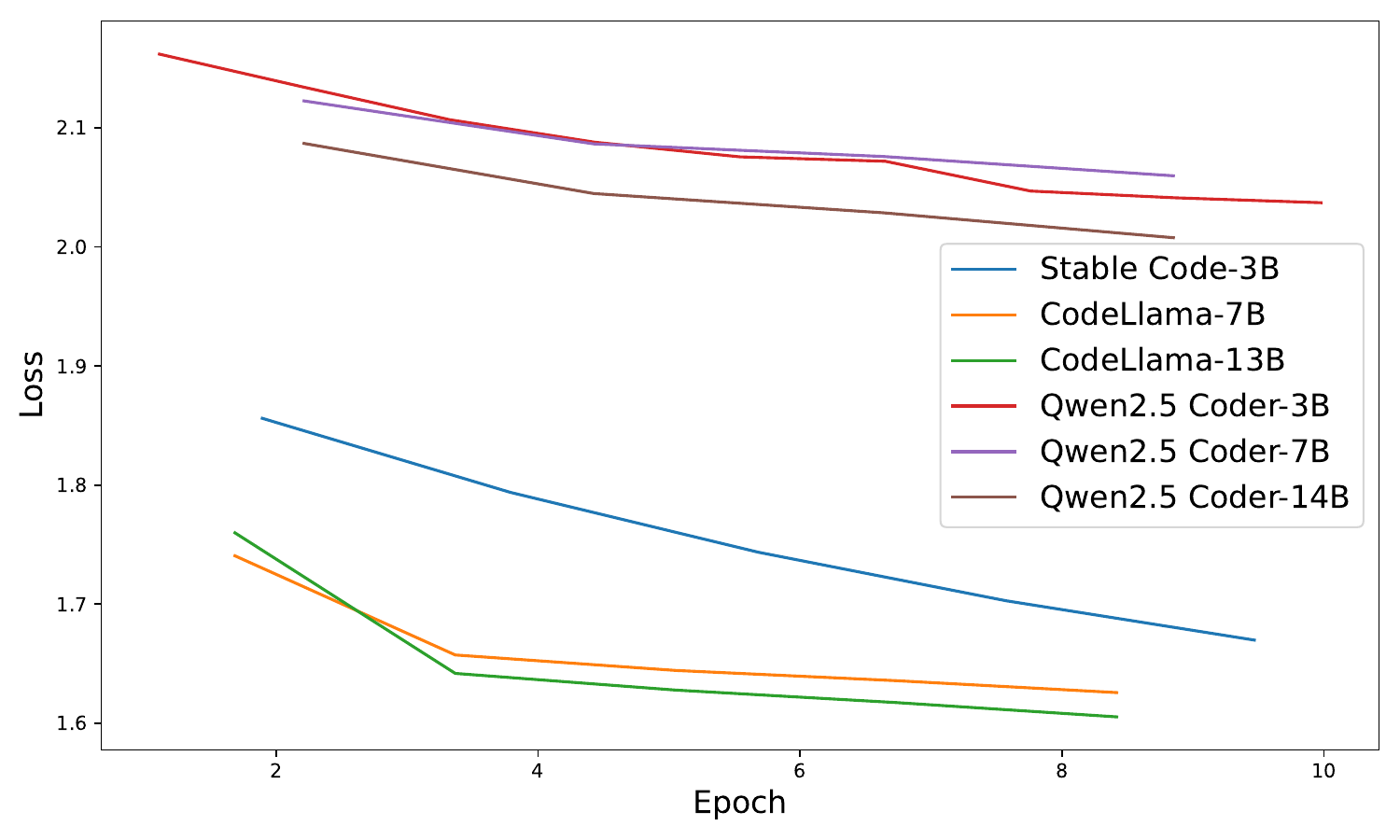}
    \caption{Training loss of the evaluated \llmsforcode models over 10
    fine-tuning epochs. The curves have different lengths because the models
    use different batch sizes.}
    \label{fig:training_loss}
  \end{minipage}
  \vspace{-.5em}
\end{figure}

\revision{We use a structural causal model (\scm) to estimate whether the learning
difficulty of a \pii instance affects its leakage during inference. This analysis is
needed because learning difficulty is observed during fine-tuning and is not assigned
through an experiment. As a result, shared factors may influence the relationship
between learning difficulty and leakage. In particular, properties of the surrounding
code may affect both how the model learns the target \pii and how easily the same value
can later be recovered from its context. The \scm in \figref{fig:scm} represents these
shared causes and supports estimation of the effect of learning difficulty after
adjusting for the observed code properties. Our unit of analysis is one \pii instance,
and all variables are defined at this level. Therefore, the analysis captures
differences between individual instances, including instances of the same \pii type.}

\revision{\textbf{Modeling causal assumptions.}
The direction of each edge in the \scm follows the temporal order of the study. First, the confounders ($Z$) describe
structural properties of a code snippet that are fixed before fine-tuning (\eg 
$<$nloc$>$, $<$token\_counts$>$, $<$\#$ast\_levels$>$, $<$\#$ast\_nodes$>$,
$<$\#$identifiers$>$, $<$\#$ast\_errors$>$, and $<$complexity$>$). These features
describe the size, nesting, and syntactic structure of the snippet. They may affect how
easily the model learns the target \pii and how much useful context is available during
an attack. We therefore model them as shared causes of the treatment and outcome
through the paths $Z \rightarrow T$ and $Z \rightarrow Y$. Second, the treatment
variable ($T$) represents the learning difficulty of the target \pii instance.
Easy-to-learn instances form the control group ($T_0$), and hard-to-learn ($T_1$) and
ambiguous-to-learn ($T_2$) instances form the two treatment groups. As described in
\secref{sec:metrics}, these groups are defined using the confidence and variability of
the model across fine-tuning epochs. Both measures are computed using only the tokens
that form the target \pii value. Easy-to-learn instances have high and stable
confidence, hard-to-learn instances have low and stable confidence, and
ambiguous-to-learn instances have confidence that varies across epochs. Finally, the
outcome variable ($Y$) records whether the \piiattack reproduces the target \pii value
from its surrounding context under the black-box threat model in
\secref{sec:threat_model}. These instance-level outcomes are then aggregated when we
report attack success rates. Learning difficulty is measured during fine-tuning, and
attack success is measured afterward. Accordingly, the graph includes the path
$T \rightarrow Y$ and excludes a reverse path from $Y$ to $T$, since a later attack
result cannot change the training dynamics observed during fine-tuning.}

\revision{\textbf{Causal analysis of token-level \pii leakage.}
The causal structure is designed for token-level analysis because the treatment and
outcome refer to the same \pii instance. Specifically, the treatment is computed from
the probabilities assigned to the target \pii tokens across fine-tuning epochs, and the
outcome records whether the same token sequence is reproduced in the model's
completion. Accordingly, the causal analysis estimates whether the learning difficulty of a \pii
instance influences its later reproduction during an attack. This distinction
is important because the surrounding code context may influence both learning and
leakage. For example, a common code pattern may make a target value easier to predict
during fine-tuning. The same pattern may also provide useful information when an
attacker constructs a prefix and suffix around the missing value. In this case,
context-based prediction contributes to the observed relationship and may be mistaken
for instance-specific memorization. The graph represents this alternative explanation
through the back-door path $T \leftarrow Z \rightarrow Y$. By adjusting for $Z$, we
compare instances with similar observed structural properties and reduce the influence
of this path. The three learning categories also match the learning behaviors that we
aim to study. Easy-to-learn instances show stable and high confidence. When an exact
\pii value is only weakly determined by its context, this behavior is more consistent
with instance-specific memorization \cite{feldman2020does,carlini2019secret}.
Hard-to-learn instances represent stable difficulty in learning the target value, and
ambiguous-to-learn instances represent unstable learning across epochs. Comparing
hard-to-learn and ambiguous-to-learn instances with easy-to-learn instances therefore
tests whether these learning patterns lead to different leakage risks.}

\revision{\textbf{Limitations.}
Our causal interpretation depends on the assumption that the selected confounders
capture the main observed properties that affect both learning difficulty and leakage.
Under this assumption, adjusting for $Z$ blocks the measured back-door path between
$T$ and $Y$, allowing us to identify the effect of learning difficulty from
observational data. The selected confounders may not capture every possible shared
cause. For example, semantic properties of a snippet may influence both token
learnability and attack success without being fully represented by size, nesting, or
syntactic metrics. Thus, unobserved confounding may remain, and the estimated effects
must be interpreted under this limitation. The graph also includes an instrument node
($I$) and an effect-modifier node ($E$), following the general causal-interpretability
template of \docode \cite{docode}. These nodes show causal roles supported by the
general framework. They do not enter our estimation because we do not claim that $I$
is a valid instrument and do not condition the estimates on $E$. We retain these nodes
in \figref{fig:scm} to show the complete template from which our graph is derived. All
reported effects are identified through adjustment for $Z$.}

\revision{\textbf{Estimating and interpreting the causal effects.}
After defining the graph and its adjustment set, we estimate two pairwise causal
effects: hard-to-learn versus easy-to-learn instances and ambiguous-to-learn versus
easy-to-learn instances. For a treatment level $t \in \{T_1,T_2\}$, the Average
Treatment Effect (\ate) is defined as
\[
\mathrm{ATE}(t,T_0)
=
\mathbb{E}[Y\mid do(T=t)]
-
\mathbb{E}[Y\mid do(T=T_0)]
\]
Here, $do(T=t)$ represents a hypothetical intervention that sets the learning category
to $t$. Thus, the \ate measures the expected difference in leakage between
hard-to-learn or ambiguous-to-learn instances and easy-to-learn instances after
adjusting for the observed structural causes \cite{Pearl2018Causality}. We identify the
causal effects using the back-door adjustment formula and estimate them with
\textit{DoWhy} \cite{dowhy}. We also examine whether the estimates depend on the thresholds used to define the learning categories. Specifically, we repeat the estimation under both settings described in \secref{sec:metrics} and report the resulting estimates with 95\% confidence intervals. In addition, we use four refutation
checks to assess the robustness of the estimates. R1 adds a random common cause, R2
replaces the treatment with a placebo treatment, R3 adds simulated unobserved
confounding related to both the treatment and outcome, and R4 repeats the estimation
on subsets of the data. Finally, we interpret each \ate according to its sign. For the
hard-to-learn versus easy-to-learn comparison, a negative \ate indicates lower expected
leakage among hard-to-learn instances, and a positive \ate indicates higher expected
leakage. For the ambiguous-to-learn versus easy-to-learn comparison, a negative \ate
indicates lower expected leakage among ambiguous-to-learn instances, and a positive
\ate indicates higher expected leakage. Therefore, the sign and size of the \ate show
how each learning category changes leakage relative to the easy-to-learn category.}

\subsection{Implementation Details}  
We fine-tune all selected \llmsforcode using LoRA \cite{hu2022lora}, a technique that efficiently updates the model weights through low-rank adaptations. LoRA, widely adopted in both academia and industry, enables efficient fine-tuning of large models by incorporating task-specific information while preserving pretrained knowledge. Our LoRA configuration uses rank $r=16$ and scaling factor $\alpha=32$, following recommendations from previous work~\cite{he2025privacyxray}.

For FIM, we set \texttt{fim\_rate} to 0.5, meaning that 50\% of training samples are formatted with the FIM objective. We employ the fine-tuning scripts provided by Hugging Face \cite{huggingface_run_fim} to ensure reproducibility. All models are fine-tuned for ten epochs to capture the training dynamics. The learning rates are set according to the model size: $2 \times 10^{-4}$ for 3B models, $1 \times 10^{-4}$ for 7-15B models. All fine-tuned models are evaluated in the code completion task with a context window of 1,024 tokens.

Because \llmsforcode are typically trained on open-source corpora, prior work has simulated attacks by incorporating the surrounding context of sensitive information collected from platforms such as GitHub into prompts~\cite{huang2024your, niu2023codexleaks}. When performing the \pii attack, we construct each code completion input by randomly sampling prefixes and suffixes surrounding a target \pii instance. \revision{Specifically, we remove the target \pii token itself along with a total of 50 neighboring tokens, split randomly between its preceding and subsequent context.}

\revision{The resulting prefix and suffix are formatted as a FIM prompt following either the PSM (Prefix-Suffix-Middle) or SPM (Suffix-Prefix-Middle) template, selected randomly for each query. No middle tokens are provided, requiring the model to generate the target \pii value from the surrounding context. \figref{fig:example} illustrates a concrete example: red-highlighted regions denote the query context provided to the model, and black-redacted regions indicate the \pii values, thereby simulating a realistic attack scenario.}

\section{Results}
\label{sec:results}

We conducted experiments to address RQs on the impact of the type of \pii on its leakage in \llmsforcode. For \ref{rq:learning_difficulty}, we evaluated the training dynamics of each \pii type during fine-tuning. For \ref{rq:leakage_relation}, we explore the relationship between training dynamics and \pii attack success risks. To answer \ref{rq:causal_analysis}, we use the causal interpretability framework to estimate the true effect of learning dynamics on \pii leakage, comparing easy, hard, and ambiguous cases across different model families.

\subsection{\ref{rq:learning_difficulty}: Learning Difficulty Across PII Types} \label{sec:learning_difficulty} To answer RQ1, we construct a dataset for each \pii type (as described in \secref{sec:dataset}) by extracting instances of that type from the original training corpus of \llmsforcode. For each \pii type, we collect 1,500 code files that contain highly sensitive \pii. We then fine-tune the multiple \llmsforcode of varying architectures and sizes in the \pii dataset. 
We report the training loss in \figref{fig:training_loss}, which shows that the fine-tuning process was effective.

    
    
\begin{figure}[]
    \centering
    \includegraphics[width=\columnwidth]{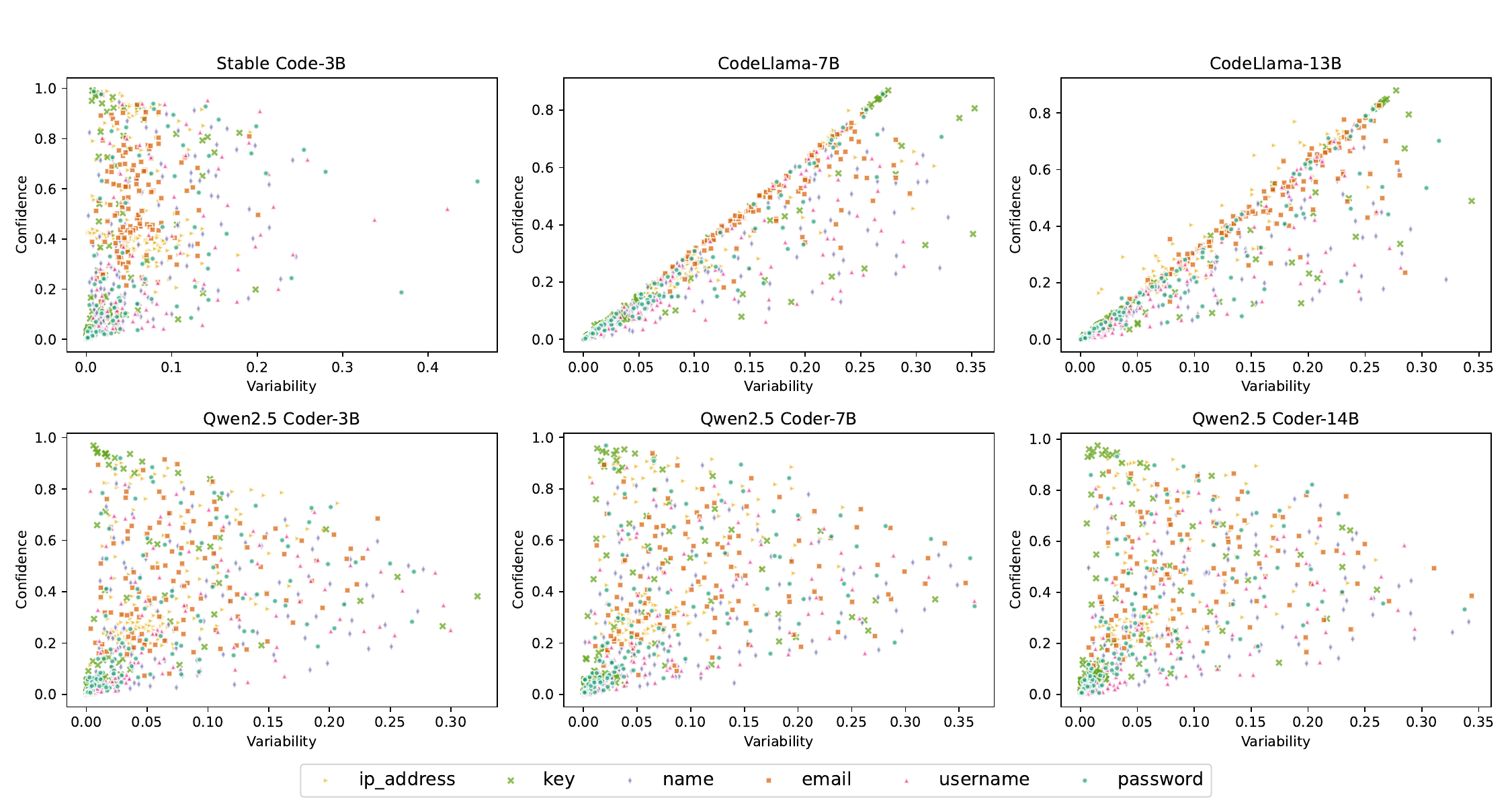}
    
    \caption{Dataset maps of the constructed training set for each fine-tuned \llmsforcode. Axes represent variability (x) and confidence (y), with samples down-sampled to preserve distribution. Unlike other models, \codeLlamaSevenB and \codeLlamaThirteenB show many high-confidence yet high-variability samples, suggesting that fine-tuning amplifies their memorization.}

    \label{fig:cartography}
    \vspace{-2em}
\end{figure}

\noindent\textbf{Qualitative analysis via Dataset Map visualization.}
To evaluate training dynamics, we measure \textit{confidence} and \textit{variability} (as defined in \secref{sec:metrics}) for each \pii instance in training epochs. We compute the training dynamics only for the tokens corresponding to \pii.
Drawing on dataset cartography~\cite{swayamdipta2020dataset}, we first provide a qualitative analysis by visualizing the training dataset for each model to understand the relative learning difficulty of different \pii types. In \figref{fig:cartography}, we illustrate the training dynamics (in the Dataset Map form~\cite{karamcheti_mind_2021}) of individual \pii instances over 10 fine-tuning epochs. The $y$-axis represents \textit{confidence}, while the $x$-axis represents \textit{variability}. Thus, the top-left corner of the data map (low variability, high confidence) corresponds to \textit{easy-to-learn} examples, the bottom-left corner (low variability, low confidence) to \textit{hard-to-learn} examples, and the right region (high variability) to \textit{ambiguous} examples~\cite{swayamdipta2020dataset}.

As illustrated in \figref{fig:cartography}, our analysis reveals distinct differences in the training dynamics of various types of \pii. Specifically, several \key and \ip instances fall within the easy-to-learn region, characterized by confidence values approaching 0.9 and variability close to 0. In contrast, a considerable number of \piiemail and \username instances occupy the ambiguous region, where the model predictions fluctuate between epochs. Furthermore, the majority of \key, along with most \password and \username instances, are categorized as hard-to-learn, consistently exhibiting low confidence values.  

These findings suggest that \ip values are comparatively easier for \llmsforcode to acquire, while \password, \username, and most \key instances remain considerably more difficult to learn. This outcome aligns with intuition: \password, \username, and \key typically consist of irregular or randomly generated character sequences that inherently lack structural regularities, making them difficult to memorize, even for humans.

\begin{figure}[h]
    \centering
    \includegraphics[width=\columnwidth]{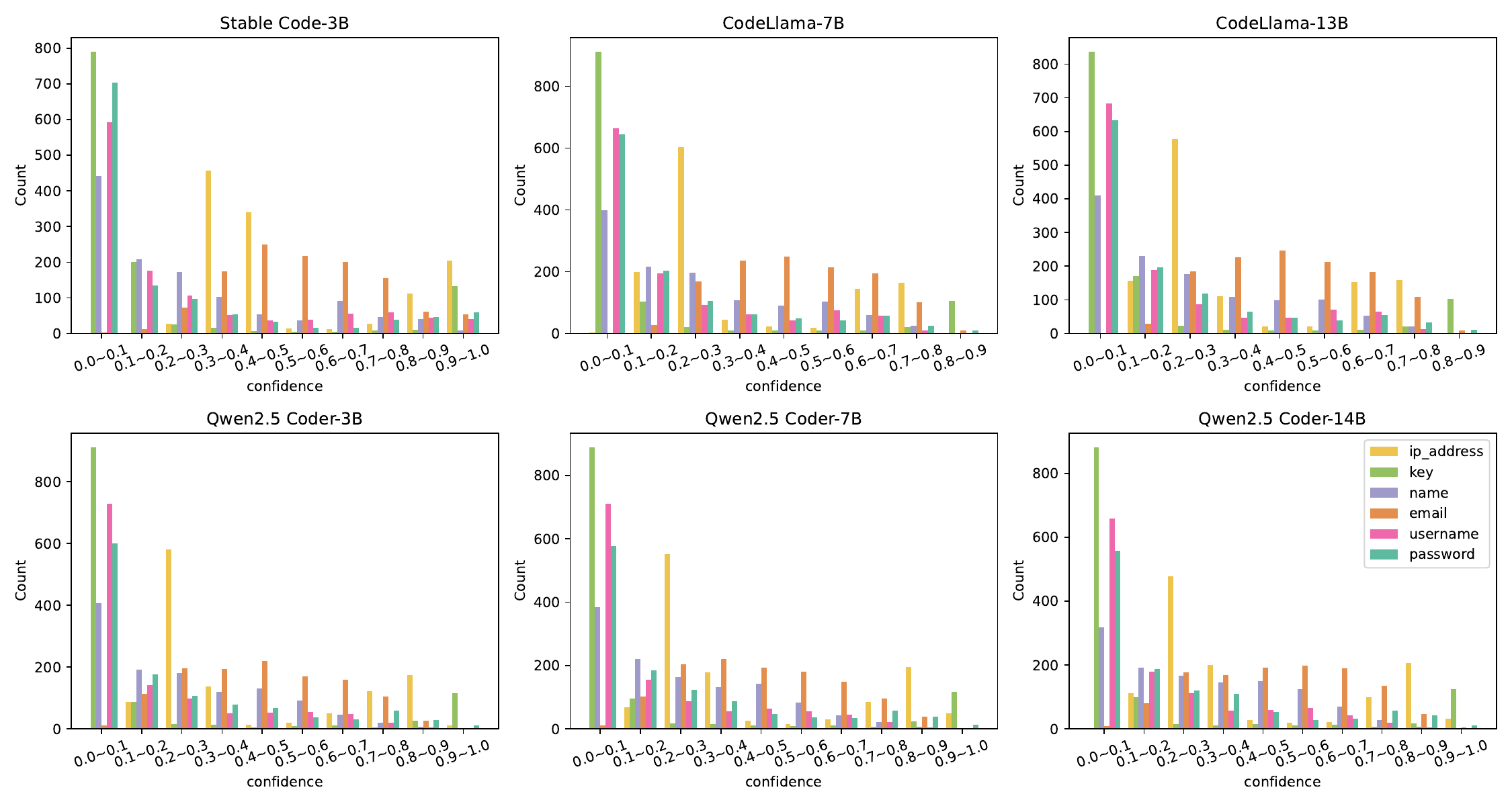}
    \vspace{-1.5em}
    \caption{Distribution of prediction confidence across \pii types for each fine-tuned \llmsforcode.}
    \label{fig:confidence}
    \vspace{-1em}
\end{figure}

\begin{figure}[h]
    \centering
    \includegraphics[width=\columnwidth]{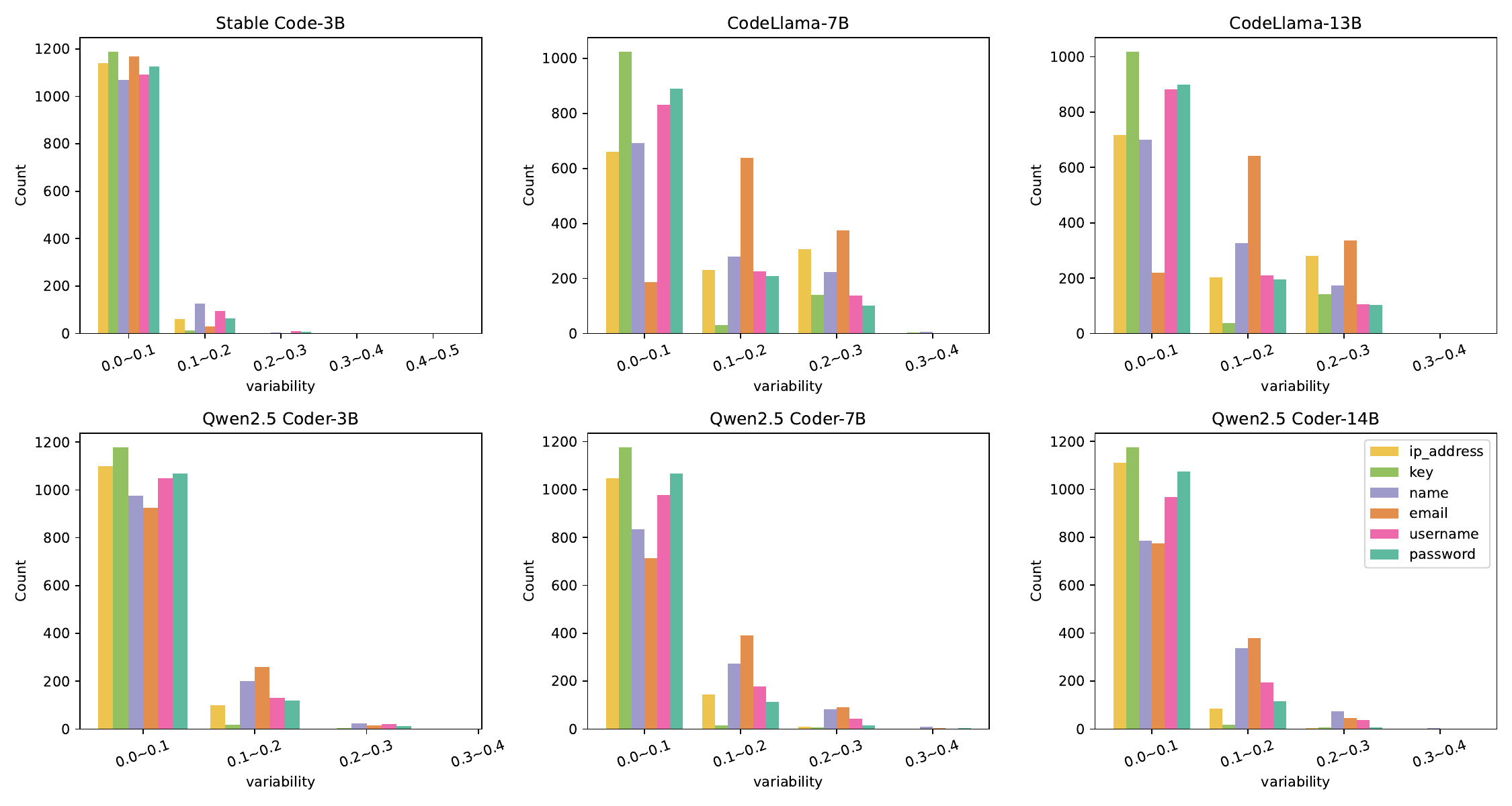}
    \vspace{-1.5em}
    \caption{Distribution of prediction variability across \pii types for each fine-tuned \llmsforcode.}
    \label{fig:variability}
    \vspace{-2em}
\end{figure}

\noindent\textbf{Quantitative analysis via metric value distribution.}
Although the Dataset Map provides qualitative insights, we also present quantitative analyses in \figref{fig:confidence} and \figref{fig:variability}. Specifically, we discretize the confidence or variability spaces into bins and compute the distribution of instances for each \pii type. \revision{The percentages below are read directly from these confidence and variability bins and therefore do not depend on the classification cutoffs introduced in \secref{sec:metrics}; where we instead refer to the easy-to-learn, ambiguous-to-learn, or hard-to-learn regions, we use the baseline 75th/25th percentile setting.} From the results across all models, several key observations emerge:

\begin{itemize}
    \item \revision{Around 13\% of \ip and \key instances consistently achieve confidence values between 0.8 and 1.0 across epochs, indicating that these examples are relatively easy for the models to learn and predict reliably.}
    \item In contrast, less than 1\% of \ip instances have confidence below 0.1, suggesting that \ip is rarely categorized as hard to learn compared to other \pii types.
    \item The most challenging instances are found among \key values: of the 1,200 examined, at least 66\% instances consistently maintain confidence below 0.1 across epochs, highlighting their inherent difficulty for model learning.
    \item \revision{A large share of \username and \password instances show low confidence values across models, which places most of them in the hard-to-learn region under the baseline 75th/25th percentile setting.}
    \item \revision{Many \piiemail, \name, and \ip instances show moderate confidence (0.1--0.5) together with relatively high variability, which places a large share of them in the ambiguous-to-learn region under the baseline 75th/25th percentile setting and indicates inconsistent model behavior across epochs.}
\end{itemize}

Taken together, these results demonstrate that different \pii types exhibit heterogeneous learning difficulty: \ip is generally easier to learn, \key shows highly polarized behavior with both easy and most hard instances, and \username and \password emerge as particularly challenging.


\begin{boxK}
\vspace{-0.5em}
    \textit{\ref{rq:learning_difficulty}}:  
    The quantitative analysis of training dynamics reveals that \ip instances are systematically easier for models to acquire, whereas the majority of \key, along with most \password and \username instances, remain considerably difficult to learn. 
\vspace{-0.5em}
\end{boxK}

\subsection{\ref{rq:leakage_relation}: Learning Difficulty–Leakage Risk Relationship} 
\label{sec:leakage_relation}

In \ref{rq:leakage_relation}, we investigate whether the training dynamics obtained in RQ1 correlate with \pii attack success rates. To this end, we perform the \pii attack. Prior studies often constructed query templates using contextual information surrounding sensitive values from GitHub~\cite{huang2024your, niu2023codexleaks}. In our setting, we have direct access to such information from the training dataset. For each \pii instance, we randomly retain part of its surrounding context while removing the target \pii itself along with 50 neighboring tokens. We then query the model with these inputs for all \pii instances, resulting in 1,200 queries per type. An attack is counted as successful if the model’s output contains the removed \pii. The results are reported in Table~\ref{tab:leakage}. \revision{To confirm that the observed leakage is introduced by fine-tuning rather than inherited from the pretrained models, we applied the identical attack protocol to the original \codeLlamaSevenB and \codeLlamaThirteenB checkpoints prior to any fine-tuning, and observed zero successful extractions across all six \pii types.}

\begin{table}[h]

\centering
\caption{Successful \pii attack instances by \llmsforcode and \pii type}
\label{tab:leakage}
\vspace{-1em}

\setlength{\tabcolsep}{4pt} 
\begin{tabular}{lcccccc}
\hline
 & \textit{\textbf{\piiemail}} & \textit{\textbf{\key}} & \textit{\textbf{\ip}} & \textit{\textbf{\name}} & \textit{\textbf{\username}} & \textit{\textbf{\password}} \\ \hline
\StableCodeThreeB         & 42       & 74 & \textbf{170} & 56           & {\ul 155}    & 90 \\
\qwenTwoFiveCoderThreeB    & {\ul 60} & 16 & 31           & 58           & \textbf{63}  & 28 \\
\codeLlamaSevenB           & 16       & 8  & 27           & \textbf{31}  & {\ul 30}     & 8  \\
\qwenTwoFiveCoderSevenB    & {\ul 93} & 26 & 68           & 85           & \textbf{117} & 48 \\
\codeLlamaThirteenB        & {\ul 26} & 5  & 13           & {\ul 26}     & \textbf{45}  & 9  \\
\qwenTwoFiveCoderFourteenB & 123      & 40 & 77           & \textbf{131} & {\ul 127}    & 52 \\ \hline
\end{tabular}
\captionsetup{justification=centering}
\caption*{\footnotesize Cells show valid reproductions. \\ \textbf{bold} = largest, \underline{underline} = second largest.}

\vspace{-2em}
\end{table}

We summarize the relationship between training dynamics and attack success rates as follows:  
\begin{itemize}
\item \key and \password instances are predominantly hard-to-learn (low confidence) in the training dynamics. Consistently, they exhibit relatively low attack success rates (e.g., for CodeLlama-7B, \key = 8 and \password = 8).  

\item \revision{\ip instances are among the easier types for the models to learn, with a notable share showing high confidence and low variability, although a considerable portion also falls into the ambiguous region. In line with their easier learning behavior, they show high attack success rates (e.g., 170 for \StableCodeThreeB).}

\item \username exhibits a remarkably high leakage count (e.g., 155 in Stable Code-3B, 117 in Qwen2.5-Coder-7B, and 127 in Qwen2.5-Coder-14B). \revision{At the same time, \username instances are rarely easy to learn, with most falling into the hard-to-learn or ambiguous categories. This trend differs from that observed for other \pii types, where high leakage tends to coincide with easy-to-learn behavior. Because \username leaks heavily despite rarely being easy to learn, a simple correlation between learning difficulty and leakage does not explain its behavior, which motivates the instance-level causal analysis in \secref{sec:rq3_results}.}

\end{itemize}

\revision{Overall, these results indicate a generally positive correlation between training dynamics and \pii attack success rates. Easy-to-learn types such as \ip are more likely to leak, whereas hard-to-learn types such as \key and \password are less prone to leakage.}
However, \username stands out as a significant exception, since although it is categorized as hard-to-learn or ambiguous in the training dynamics, it consistently exhibits high leakage across different models. To better understand this discrepancy, we further conduct an instance-level causal analysis to investigate the relationship between training dynamics and attack success rates.
\begin{boxK}
\vspace{-0.5em}
    \textit{\ref{rq:leakage_relation}}:
\revision{In the correlation analysis, easy-to-learn types such as \ip exhibit high leakage risks under attack, whereas hard-to-learn types such as \key and \password show low leakage. However, \username emerges as a significant exception, since despite being mostly classified as hard-to-learn or ambiguous, it consistently demonstrates high leakage across models, which warrants further causal analysis.}
\vspace{-0.5em}
\end{boxK}

\subsection{\ref{rq:causal_analysis}: Causal Effect of Learning Dynamics on Leakage Risk}
\label{sec:rq3_results}
\begin{table}[]

\centering
\caption{Causal Analysis Results (Easy-to-Learn vs Hard-to-Learn)}
\label{tab:causal_results_hard}
\vspace{-1em}
\scalebox{0.65}{

\setlength{\tabcolsep}{4pt} 
\begin{tabular}{lllllllllllll}
 &
  \multicolumn{2}{c}{\StableCodeThreeB} &
  \multicolumn{2}{c}{\qwenTwoFiveCoderThreeB} &
  \multicolumn{2}{c}{\codeLlamaSevenB} &
  \multicolumn{2}{c}{\qwenTwoFiveCoderSevenB} &
  \multicolumn{2}{c}{\codeLlamaThirteenB} &
  \multicolumn{2}{c}{\qwenTwoFiveCoderFourteenB} \\ \hline
\textbf{\pii Type} &
  \textbf{\correlation} &
  \textbf{\ate} &
  \textbf{\correlation} &
  \textbf{\ate} &
  \textbf{\correlation} &
  \textbf{\ate} &
  \textbf{\correlation} &
  \textbf{\ate} &
  \textbf{\correlation} &
  \textbf{\ate} &
  \textbf{\correlation} &
  \textbf{\ate} \\ \hline
\textit{name} &
  \textbf{-0.06 $\pm$ 0.05} &
  \cellcolor[HTML]{9E9BCA}-0.03 $\pm$ 0.01 &
  \textbf{-0.10 $\pm$ 0.03} &
  \cellcolor[HTML]{9E9BCA}-0.03 $\pm$ 0.02 &
  0.00 $\pm$ 0.00 &
  0.00 $\pm$ 0.00 &
  -0.03 $\pm$ 0.07 &
  \cellcolor[HTML]{9E9BCA}-0.03 $\pm$ 0.03 &
  -0.01 $\pm$ 0.21 &
  0.00 $\pm$ 0.04 &
  -0.02 $\pm$ 0.04 &
  -0.01 $\pm$ 0.02 \\
\textit{key} &
  -0.02 $\pm$ 0.05 &
  -0.09 $\pm$ 0.20 &
  -0.11 $\pm$ 0.12 &
  \cellcolor[HTML]{9E9BCA}-0.03 $\pm$ 0.01 &
  0.00 $\pm$ 0.00 &
  0.00 $\pm$ 0.00 &
  -0.12 $\pm$ 0.14 &
  -0.05 $\pm$ 0.05 &
  0.00 $\pm$ 0.00 &
  0.00 $\pm$ 0.00 &
  \textbf{-0.20 $\pm$ 0.17} &
  \cellcolor[HTML]{9E9BCA}-0.10 $\pm$ 0.07 \\
\textit{username} &
  \textbf{-0.63 $\pm$ 0.03} &
  \cellcolor[HTML]{9E9BCA}-0.46 $\pm$ 0.08 &
  \textbf{-0.32 $\pm$ 0.05} &
  \cellcolor[HTML]{9E9BCA}-0.09 $\pm$ 0.01 &
  0.00 $\pm$ 0.00 &
  0.00 $\pm$ 0.00 &
  \textbf{-0.34 $\pm$ 0.02} &
  \cellcolor[HTML]{9E9BCA}-0.14 $\pm$ 0.01 &
  0.00 $\pm$ 0.00 &
  0.00 $\pm$ 0.00 &
  \textbf{-0.35 $\pm$ 0.01} &
  \cellcolor[HTML]{9E9BCA}-0.15 $\pm$ 0.04 \\
\textit{password} &
  \textbf{-0.15 $\pm$ 0.14} &
  -0.03 $\pm$ 0.04 &
  \textbf{-0.10 $\pm$ 0.02} &
  \cellcolor[HTML]{9E9BCA}-0.03 $\pm$ 0.01 &
  0.01 $\pm$ 0.01 &
  0.00 $\pm$ 0.00 &
  \textbf{-0.08 $\pm$ 0.05} &
  \cellcolor[HTML]{9E9BCA}-0.03 $\pm$ 0.01 &
  {\ul \textbf{0.01 $\pm$ 0.00}} &
  \cellcolor[HTML]{C0C0C0}0.00 $\pm$ 0.00 &
  \textbf{-0.10 $\pm$ 0.07} &
  -0.03 $\pm$ 0.04 \\
\textit{email} &
  -0.39 $\pm$ 0.09 &
  \cellcolor[HTML]{9E9BCA}-0.20 $\pm$ 0.10 &
  -0.12 $\pm$ 0.01 &
  -0.02 $\pm$ 0.04 &
  0.00 $\pm$ 0.00 &
  0.00 $\pm$ 0.00 &
  -0.17 $\pm$ 0.05 &
  \cellcolor[HTML]{9E9BCA}-0.03 $\pm$ 0.03 &
  0.00 $\pm$ 0.00 &
  0.00 $\pm$ 0.00 &
  -0.16 $\pm$ 0.15 &
  -0.03 $\pm$ 0.05 \\
\textit{ip\_address} &
  \textbf{-0.42 $\pm$ 0.10} &
  \cellcolor[HTML]{9E9BCA}-0.20 $\pm$ 0.16 &
  \textbf{-0.22 $\pm$ 0.11} &
  \cellcolor[HTML]{9E9BCA}-0.24 $\pm$ 0.15 &
  {\ul \textbf{0.01 $\pm$ 0.00}} &
  0.00 $\pm$ 0.00 &
  \textbf{-0.05 $\pm$ 0.02} &
  -0.06 $\pm$ 0.06 &
  {\ul \textbf{0.01 $\pm$ 0.01}} &
  \cellcolor[HTML]{C0C0C0}0.01 $\pm$ 0.00 &
  \textbf{-0.06 $\pm$ 0.04} &
  -0.01 $\pm$ 0.04 \\ \hline
\end{tabular}
} 
\captionsetup{justification=centering}
\caption*{\footnotesize \small{\revision{Values are 95\% confidence intervals of the computed scores across the two cutoff settings in \secref{sec:metrics}}. bold: \textnormal{$-$ \correlation}, \underline{bold underlined}: \textnormal{$+$ \correlation}, {\color[HTML]{9e9bca} background purple}: \textnormal{$-$ causal effect}, {\color[HTML]{c4c3c4} background grey}: \textnormal{$+$ causal effect}}}

\vspace{-2em}
\end{table}

Having analyzed learning difficulty in \secref{sec:learning_difficulty} and its correlation with leakage in \secref{sec:leakage_relation}, we now turn to causal inference in \ref{rq:causal_analysis}. The goal is to move beyond correlation and determine whether training dynamics themselves exert a causal effect on leakage risk. To do so, we estimate the Average Treatment Effect (\ate) when comparing easy-to-learn instances against those that are hard or ambiguous. \ates allow us to test whether differences in learning dynamics directly shape the likelihood of \pii leakage.

\begin{table*}[]

\centering
\caption{Causal Analysis Results (Easy vs Ambiguous to Learn)}
\label{tab:causal_results_ambiguous}
\vspace{-1em}
\scalebox{0.65}{

\setlength{\tabcolsep}{4pt} 
\begin{tabular}{lllllllllllll}
 &
  \multicolumn{2}{c}{\StableCodeThreeB} &
  \multicolumn{2}{c}{\qwenTwoFiveCoderThreeB} &
  \multicolumn{2}{c}{\codeLlamaSevenB} &
  \multicolumn{2}{c}{\qwenTwoFiveCoderSevenB} &
  \multicolumn{2}{c}{\codeLlamaThirteenB} &
  \multicolumn{2}{c}{\qwenTwoFiveCoderFourteenB} \\ \hline
\textbf{\pii Type} &
  \textbf{\correlation} &
  \textbf{\ate} &
  \textbf{\correlation} &
  \textbf{\ate} &
  \textbf{\correlation} &
  \textbf{\ate} &
  \textbf{\correlation} &
  \textbf{\ate} &
  \textbf{\correlation} &
  \textbf{\ate} &
  \textbf{\correlation} &
  \textbf{\ate} \\ \hline
\textit{name} &
  0.01 $\pm$ 0.02 &
  0.00 $\pm$ 0.01 &
  {\ul \textbf{0.04 $\pm$ 0.02}} &
  \cellcolor[HTML]{C0C0C0}0.03 $\pm$ 0.00 &
  {\ul \textbf{0.03 $\pm$ 0.02}} &
  \cellcolor[HTML]{C0C0C0}0.01 $\pm$ 0.00 &
  {\ul \textbf{0.05 $\pm$ 0.03}} &
  0.04 $\pm$ 0.06 &
  0.02 $\pm$ 0.06 &
  0.01 $\pm$ 0.04 &
  {\ul \textbf{0.08 $\pm$ 0.01}} &
  \cellcolor[HTML]{C0C0C0}0.07 $\pm$ 0.02 \\
\textit{key} &
  0.01 $\pm$ 0.03 &
  0.02 $\pm$ 0.04 &
  -0.03 $\pm$ 0.06 &
  -0.02 $\pm$ 0.04 &
  {\ul \textbf{0.07 $\pm$ 0.03}} &
  \cellcolor[HTML]{C0C0C0}0.03 $\pm$ 0.02 &
  {\ul \textbf{0.05 $\pm$ 0.04}} &
  \cellcolor[HTML]{C0C0C0}0.03 $\pm$ 0.03 &
  {\ul \textbf{0.06 $\pm$ 0.01}} &
  \cellcolor[HTML]{C0C0C0}0.01 $\pm$ 0.01 &
  0.00 $\pm$ 0.09 &
  -0.01 $\pm$ 0.08 \\
\textit{username} &
  {\ul \textbf{-0.32 $\pm$ 0.00}} &
  \cellcolor[HTML]{9E9BCA}-0.23 $\pm$ 0.10 &
  -0.16 $\pm$ 0.16 &
  \cellcolor[HTML]{9E9BCA}-0.10 $\pm$ 0.07 &
  0.02 $\pm$ 0.04 &
  0.02 $\pm$ 0.04 &
  {\ul \textbf{-0.07 $\pm$ 0.04}} &
  -0.05 $\pm$ 0.15 &
  0.02 $\pm$ 0.03 &
  \cellcolor[HTML]{C0C0C0}0.02 $\pm$ 0.00 &
  {\ul \textbf{0.06 $\pm$ 0.05}} &
  0.05 $\pm$ 0.06 \\
\textit{password} &
  {\ul \textbf{0.11 $\pm$ 0.07}} &
  \cellcolor[HTML]{C0C0C0}0.08 $\pm$ 0.03 &
  {\ul \textbf{0.04 $\pm$ 0.00}} &
  0.03 $\pm$ 0.05 &
  {\ul \textbf{0.06 $\pm$ 0.02}} &
  \cellcolor[HTML]{C0C0C0}0.05 $\pm$ 0.01 &
  {\ul \textbf{0.06 $\pm$ 0.02}} &
  \cellcolor[HTML]{C0C0C0}0.07 $\pm$ 0.02 &
  {\ul \textbf{0.04 $\pm$ 0.01}} &
  \cellcolor[HTML]{C0C0C0}0.05 $\pm$ 0.01 &
  {\ul \textbf{0.10 $\pm$ 0.01}} &
  \cellcolor[HTML]{C0C0C0}0.11 $\pm$ 0.06 \\
\textit{email} &
  0.04 $\pm$ 0.06 &
  0.06 $\pm$ 0.07 &
  {\ul \textbf{0.07 $\pm$ 0.01}} &
  \cellcolor[HTML]{C0C0C0}0.04 $\pm$ 0.00 &
  {\ul \textbf{0.03 $\pm$ 0.01}} &
  \cellcolor[HTML]{C0C0C0}0.03 $\pm$ 0.01 &
  {\ul \textbf{0.12 $\pm$ 0.01}} &
  \cellcolor[HTML]{C0C0C0}0.08 $\pm$ 0.02 &
  {\ul \textbf{0.03 $\pm$ 0.01}} &
  \cellcolor[HTML]{C0C0C0}0.03 $\pm$ 0.00 &
  {\ul \textbf{0.13 $\pm$ 0.05}} &
  \cellcolor[HTML]{C0C0C0}0.08 $\pm$ 0.07 \\
\textit{ip\_address} &
  -0.01 $\pm$ 0.08 &
  0.03 $\pm$ 0.02 &
  -0.05 $\pm$ 0.08 &
  -0.17 $\pm$ 0.09 &
  {\ul \textbf{0.04 $\pm$ 0.01}} &
  0.06 $\pm$ 0.01 &
  {\ul \textbf{0.06 $\pm$ 0.01}} &
  0.09 $\pm$ 0.10 &
  {\ul \textbf{0.05 $\pm$ 0.03}} &
  0.11 $\pm$ 0.03 &
  0.04 $\pm$ 0.04 &
  0.10 $\pm$ 0.06 \\ \hline
\end{tabular}
} 
\captionsetup{justification=centering}
\caption*{\footnotesize \small{\revision{Values are 95\% confidence intervals of the computed scores across the two cutoff settings in \secref{sec:metrics}}. bold: \textnormal{$-$ \correlation}, \underline{bold underlined}: \textnormal{$+$ \correlation}, {\color[HTML]{9e9bca} background purple}: \textnormal{$-$ causal effect}, {\color[HTML]{c4c3c4} background grey}: \textnormal{$+$ causal effect}}}
\vspace{-4mm}
\end{table*}

\revision{\tabref{tab:causal_results_hard} shows the comparison between easy-to-learn and hard-to-learn instances. The \username yields the clearest pattern, with consistently negative \ates wherever an effect is estimated, and the largest magnitude of any type (for example, $-0.46$ on \StableCodeThreeB). This means that hard \username instances leak less than easy ones, so \username is more likely to be leaked when it is easier to learn. The \ip also shows strong negative \ates in some models (for example, $-0.24$ on \qwenTwoFiveCoderThreeB), but the effect is near zero in others, so ease of learning explains only part of its leakage behavior. The \key, \name, \password, and \piiemail all lean negative as well, but their \ates are small, so for these types the link between ease of learning and leakage is weak. We also note that the \codeLlamaSevenB and \codeLlamaThirteenB columns are zero across all types. This is because these two models did not produce enough instances in both the easy and hard categories to form a valid contrast, so no \ate could be estimated for them, and these zeros should not be read as evidence of no effect.}


\revision{\tabref{tab:causal_results_ambiguous} compares easy-to-learn and ambiguous-to-learn instances. The \username again stands out, but with a pattern that depends on model scale. In smaller models such as \StableCodeThreeB and \qwenTwoFiveCoderThreeB the \ates are negative, so ambiguity reduces leakage, while in larger models such as \qwenTwoFiveCoderFourteenB they turn slightly positive, so ambiguity mildly increases leakage. This scale dependence helps explain why \username leaked heavily even though it was rarely easy to learn. The \password and \piiemail show small but consistently positive \ates across models, meaning that ambiguous cases leak more often than easy ones. The \name follows the same direction, with positive \ates that grow in the larger models. This suggests that when a model is uncertain about an instance, it may still memorize fragments or fall back on memorized patterns, which can increase exposure rather than reduce it. The \key shows small and mixed \ates with no consistent sign, so ambiguity has little clear effect on it. The \ip is mostly positive but unstable, with one large negative value on \qwenTwoFiveCoderThreeB, so its response to ambiguity is not consistent.}

\revision{Taken together, the causal analysis shows that learning dynamics influence leakage in type-specific and model-dependent ways. Ease of learning most clearly drives leakage for the \username, and to a weaker extent for the \ip. Ambiguity plays a more complex role, since it reduces \username leakage in small models but mildly increases it in large ones, and it tends to increase leakage for the \name, \password, and \piiemail.}

\begin{boxK}
\vspace{-0.5em}
   \revision{\textit{\ref{rq:causal_analysis}}: \revision{Learning dynamics causally shape \pii leakage risk, and the effect depends on both \pii type and model. The \username leaks more when it is easier to learn, and this is the strongest and most consistent causal effect we observe. For the \username, ambiguity lowers leakage in smaller models but mildly raises it in larger ones. Ambiguous instances tend to leak more than easy ones for the \name, \password, and \piiemail. The remaining types show weak or unstable effects.}}
\vspace{-0.5em}
\end{boxK}

\section{Discussion}
\label{sec:discussion}

\subsection{Findings and Implications}

\noindent\textbf{LLM Refinement with proper instruction can be an effective step to reduce false positives.} 
A common challenge in \pii detection is reducing false positives—for example, \password in test files are often non-sensitive, yet most tools cannot systematically exclude them. We demonstrate that carefully designed prompts can mitigate this issue. By instructing the LLM with rules such as “\textit{If the VALUE is used in a test case, such as in a Test function or @Test annotation, it should be given a low score.}” The refinement process eliminates the vast majority of test-related false positives. In our experiments, only 9.09\% of \password instances met the criteria for sensitive \pii, underscoring the effectiveness of our LLM refinement.


\noindent\revision{\textbf{Fine-grained, type-aware privacy risk assessment.}} Our study demonstrates that privacy risks are not uniform across \pii types. Instead, leakage varies with both type and learning dynamics. \revision{For example, \username exhibits substantially higher leakage when it is easy to learn, which is the strongest causal effect we observe, whereas \name, \password, and \piiemail sometimes leak more in ambiguous cases for larger models. By contrast, \ip shows high leakage in the correlation analysis but limited sensitivity to the learning-difficulty treatment in the causal analysis, suggesting that factors other than the measured learning-difficulty category may contribute to its exposure.} Our results indicate that mitigation cannot rely on a single defense strategy but should account for the distinct risk profiles of different \pii types.

\noindent\revision{\textbf{Learnability-guided prioritization of detection effort.}} Our findings indicate that different types of \pii exhibit varying levels of learnability for \llmsforcode, suggesting that defenses should be type-specific rather than uniform. \revision{The training-dynamics statistics used in RQ1 provide one practical signal for prioritizing limited review resources. For example, a maintainer could monitor per-token confidence and variability across fine-tuning checkpoints and prioritize detected \pii instances in the easy-to-learn region for pre-release review. Because our causal analysis finds the strongest positive effect for easy-to-learn \username instances, such usernames could be prioritized for targeted data cleaning or machine unlearning~\cite{geng2025mitigating}.}

\noindent\revision{\textbf{Severity- and context-aware validation.}}\revision{Severity provides a complementary axis for prioritization. Although the estimated causal effect for \ip is limited, an exposed address may reveal an internal or cloud-hosted infrastructure endpoint. A lightweight validator could therefore combine address properties with code context. For example, private addresses appearing near database-connection or deployment code could be flagged as potential internal endpoints, while public addresses could be checked for routability and cross-referenced against published cloud-provider ranges. This design is consistent with prior work advocating context- and policy-aware secret scanning~\cite{huang2024your}.}

\noindent\revision{\textbf{Type-specific privacy protection.} Removal alone may be insufficient because PII detectors for code can be noisy. For example, only 9.09\% of the pre-checked Password candidates in our pipeline passed the LLM-based sensitivity criterion, illustrating the difficulty of applying a broad removal policy without discarding substantial amounts of benign code. Our findings suggest that additional protection should depend on both PII type and learning dynamics. For easy-to-learn \username instances, targeted data cleaning or machine unlearning may be appropriate. For \name and \password, however, ambiguous instances can exhibit greater leakage in larger models. These instances could instead be prioritized for stronger differential-privacy protection~\cite{kulynych2024attack} or targeted machine unlearning~\cite{geng2025mitigating}. These examples are preliminary design directions rather than evaluated defenses; future work should measure their effects on both privacy leakage and code-generation utility.}

\noindent\revision{\textbf{Transferability to real-world deployed systems.} In practice, \llmsforcode such as GitHub Copilot and Codex are augmented with post-training alignment (e.g., RLHF) and inference-time PII filters. While our study does not directly evaluate these mechanisms, prior work demonstrates that such defenses attenuate but do not eliminate memorization-based leakage: Ippolito et al.~\cite{ippolito2023preventing} show that verbatim-blocking filters are bypassable via style-transfer prompts, and Huang et al.~\cite{huang2024your} extracted 2,702 format-filtered credential-like outputs from Copilot despite its production safety layers. Because the memorization we characterize is encoded in model weights during training, it persists beneath inference-time defenses as a latent risk.} \revision{Our findings suggest that developers should concentrate alignment effort on the \pii types that our causal analysis identifies as most exposed, such as \username, rather than applying uniform defense strength, since overly aggressive uniform removal is known to degrade model utility~\cite{jang2023knowledge}.}

\subsection{Threats to Validity}
\label{sec:threats_to_validity}

We first discuss \textbf{the internal validity of our study}, i.e., the correctness of the experimental implementation. To ensure the reliability of the collected \pii dataset, we recruited two experienced researchers in software engineering to perform manual verification, as described in \secref{expert}. At a 95\% confidence level and a 5\% margin of error, we estimate that 84.2\% to 94.5\% of the dataset consists of genuine sensitive \pii instances. For fine-tuning, we reused the official Hugging Face implementation for FIM-based fine-tuning\footnote{\url{https://github.com/huggingface/transformers/blob/main/examples/pytorch/language-modeling/run_fim.py}} and verified that all models employed in our experiments natively support the FIM objective.
\textbf{Threats to external validity} arise from the representativeness of the studied \pii types. To address this concern, we extracted \pii directly from real-world training corpora to ensure authenticity. 
\revision{Our study focuses exclusively on Java, which introduces two types of potential generalizability concerns. First, the absolute leakage rates in~\tabref{tab:leakage} may partly reflect Java-specific conventions. For example, JDBC connection strings and Spring framework annotations often appear in templated contexts surrounding PII tokens, which may amplify their learnability compared to other languages. 
Second, our causal finding that easier-to-learn PII instances are more susceptible to extraction is grounded in general properties of LLM training processes that are not language-specific~\cite{carlini2022quantifying}. Moreover, many PII types, such as email addresses and IP addresses, follow standardized lexical formats (e.g., RFC 5321, RFC 791), whose structural regularity is independent of the surrounding programming language. 
We therefore expect the relative leakage patterns and underlying mechanisms to generalize across languages, although the exact leakage rates may vary and require empirical validation. Extending this study to other languages, such as Python and JavaScript, is left as future work.
}
Another external validity concern is the size of our dataset, which is relatively small compared to the massive corpora used for pre-training \llmsforcode. Although our dataset is representative of common fine-tuning scenarios, it may not capture the full range of memorization behaviors that emerge in large-scale training regimes with millions or billions of samples. This limitation introduces uncertainty in generalizing our results to production-level fine-tuning.  
Finally, \textbf{threats to construct validity} refer to the suitability of our evaluation. In this study, we evaluated fine-tuning effectiveness using code completion token accuracy. For privacy leakage, we simulated realistic \pii attack strategies and measured the \pii leakage rate, which is consistent with prior research~\cite{huang2024your, niu2023codexleaks}. \revision{A further construct validity concern is the discretization of training dynamics into easy, ambiguous, and hard categories, since the quantile cutoffs used for this step are a researcher choice that could influence the resulting categories and the causal estimates derived from them. To mitigate this concern, we compute all causal estimates and correlations under the baseline 75th/25th percentile setting and the stricter 80th/20th percentile setting described in \secref{sec:metrics}. The conclusions remain stable under the stricter setting, indicating that the findings are not driven by the baseline cutoff choice. The causal conclusions remain stable across these sets, which indicates that our findings do not depend on one specific way of splitting learning difficulty.} Together, these evaluations provide a sound methodological basis for assessing the learning and leakage of different \pii types in \llmsforcode.

\section{Conclusion}
\label{sec:conclusions}
This paper introduced a causality-based approach to interpret privacy risks in \llmsforcode by linking training dynamics with \pii leakage behavior. Through a carefully constructed multi-type \pii dataset, fine-tuning experiments, and structural causal modeling, we showed that leakage risks are not uniform but shaped by both the type of \pii and its learnability. \revision{In the correlation analysis, easy-to-learn cases such as \textit{IP addresses} tend to leak readily, while hard-to-learn cases like \textit{keys} and \textit{passwords} leak less often. The causal analysis refines this picture, showing that ease of learning has a clear causal effect on leakage for \textit{usernames}, the strongest and most consistent effect we observe, while the high leakage of \textit{IP addresses} is not matched by a comparable causal effect. Ambiguous cases exhibit more nuanced, model-dependent behavior, reducing leakage in some settings while increasing it in others.} By establishing causal evidence for these patterns, our work provides actionable insights for designing type- and learning-aware defenses, highlighting that effective safeguards must move beyond one-size-fits-all strategies to account for the heterogeneous privacy risks inherent in code models.


\bibliographystyle{ACM-Reference-Format}
\bibliography{ref}

\end{document}